\documentclass[twocolumn]{aastex62}

\usepackage{bm}

\received{February 13, 2020}
\revised{August 25, 2020}
\accepted{August 27,2020}
\submitjournal{ApJ}

\shorttitle{Kinetic Instabilities In The Solar Wind}
\shortauthors{Sun et al.}

\begin{document}

\title{Electron Temperature Anisotropy and Electron Beam Constraints From Electron Kinetic Instabilities in the Solar Wind}

\correspondingauthor{Jinsong Zhao}
\email{js\_zhao@pmo.ac.cn}

\author{Heyu Sun}
\affil{Key Laboratory of Planetary Sciences, Purple Mountain Observatory, Chinese Academy of Sciences, Nanjing 210008, People's Republic of China}
\affil{School of Astronomy and Space Science, University of Science and Technology of China, Hefei 230026, People's Republic of China}

\author{Jinsong Zhao}
\affiliation{Key Laboratory of Planetary Sciences, Purple Mountain Observatory, Chinese Academy of Sciences, Nanjing 210008, People's Republic of China}
\affil{School of Astronomy and Space Science, University of Science and Technology of China, Hefei 230026, People's Republic of China}

\author{Wen Liu}
\affil{Key Laboratory of Planetary Sciences, Purple Mountain Observatory, Chinese Academy of Sciences, Nanjing 210008, People's Republic of China}
\affil{School of Astronomy and Space Science, University of Science and Technology of China, Hefei 230026, People's Republic of China}

\author{Huasheng Xie}
\affiliation{Hebei Key Laboratory of Compact Fusion, Langfang 065001, People's Republic of China}
\affiliation{ENN Science and Technology Development Co., Ltd., Langfang 065001, People's Republic of China}

\author{Dejin Wu}
\affiliation{Key Laboratory of Planetary Sciences, Purple Mountain Observatory, Chinese Academy of Sciences, Nanjing 210008, People's Republic of China}

\begin{abstract}
Electron temperature anisotropies and electron beams are nonthermal features of the observed nonequilibrium electron velocity distributions in the solar wind. 
In collision-poor plasmas these nonequilibrium distributions are expected to be regulated by kinetic instabilities through wave-particle interactions.
This study considers electron instabilities driven by the interplay of core electron temperature anisotropies and the electron beam, and firstly gives a comprehensive analysis of instabilities in arbitrary directions to the background magnetic field.
It clarifies the dominant parameter regime (e.g., parallel core electron plasma beta $\beta_{\mathrm{ec\parallel}}$, core electron temperature anisotropy $A_{\mathrm{ec}}\equiv T_{\mathrm{ec\perp}}/T_{\mathrm{ec\parallel}}$, and electron beam velocity $V_{\mathrm{eb}}$) for each kind of electron instability (e.g., the electron beam-driven electron acoustic/magnetoacoustic instability, the electron beam-driven whistler instability, the electromagnetic electron cyclotron instability, the electron mirror instability, the electron firehose instability, and the ordinary-mode instability). It finds that the electron beam can destabilize electron acoustic/magnetoacoustic waves in the low-$\beta_{\mathrm{ec\parallel}}$ regime, and whistler waves in the medium- and large-$\beta_{\mathrm{ec\parallel}}$ regime. It also finds that a new oblique fast-magnetosonic/whistler instability is driven by the electron beam with $V_{\mathrm{eb}}\gtrsim7V_{\mathrm{A}}$ in a regime where $\beta_{\mathrm{ec\parallel}}\sim0.1-2$ and $A_{\mathrm{ec}}<1$.
Moreover, this study presents electromagnetic responses of each kind of electron instability.
These results provide a comprehensive overview for electron instability constraints on core electron temperature anisotropies and electron beams in the solar wind.

\end{abstract}

\keywords{plasmas --- instabilities --- solar wind}

\section{Introduction} \label{sec:intro}

In situ measurements of the electron velocity distribution functions (eVDFs) in the solar wind reveal states out of thermal equilibrium, including the temperature anisotropy; that is, the temperature $(T_{\mathrm{e\perp}})$ perpendicular to the ambient magnetic field is different from the temperature $(T_{\mathrm{e\parallel}})$ parallel to the ambient magnetic field \citep[e.g.,][]{1975JGR....80.4181F,1987JGR....92.1075P,2008JGRA..113.3103S,2012ApJ...753L..23W,2016SoPh..291.2165P}. Large temperature anisotropies can trigger different kinetic instabilities, which enhance the electromagnetic fluctuations. These fluctuations interact with electrons, and in turn, the electron anisotropic velocity distribution tends to reach a the quasi-stable state. Therefore, electron kinetic instabilities may provide constraints on electron temperatures in the solar wind \citep{2008JGRA..113.3103S,2016ApJ...825L..26C,2017MNRAS.464..564L,2019A&A...627A..76S,2019ApJ...871..237S}. 

Different electron kinetic instabilities are driven by different kinds of electron temperature anisotropies, i.e., the perpendicular temperature anisotropy ($A_e>1$) and the parallel temperature anisotropy ($A_e<1$), where $A_e\equiv T_{\mathrm{e\perp}}/T_{\mathrm{e\parallel}}$. The $A_e>1$ can induce the electromagnetic electron cyclotron instability \citep[or the whistler instability; e.g.,][]{1966JGR....71....1K,1996JGR...10110749G,2017A&A...602A..44L,2018JGRA..123....6L,2019Ap&SS.364..171L,2019A&A...627A..76S,2019ApJ...883..185Z} and the electron mirror instability \citep[e.g.,][]{
2006JGRA..11111224G,2018JPlPh..84d9002H,2018JGRA..123.1754S}. The $A_e<1$ can drive the periodic electron firehose instability \citep[or the parallel electron firehose instability; e.g.,][]{1970JGR....75.5297H,1985JGR....90.7607G,2017MNRAS.464..564L,2017PhPl...24a2907S,2017PhPl...24k2104Y,2019ApJ...871..237S}, the aperiodic electron firehose instability \citep[or the oblique electron firehose instability; e.g.,][] {1999A&A...351..741P,2000JGR...10527377L,2003PhPl...10.3571G,2008JGRA..113.7107C,2010ApJ...710.1848C,2014JGRA..119...59H,2019ApJ...873L..20L,2019MNRAS.483.5642S}, and the ordinary-mode instability \citep[e.g.,][]{1970PhFl...13.1407D,2012PhPl...19g2116I,2014SoPh..289..369L,2015PhPl...22h2122S}. Since the electromagnetic electron cyclotron instability (the aperiodic electron firehose instability) is generally stronger than the electron mirror instability (the periodic electron firehose instability) at the same plasma condition, electromagnetic electron cyclotron and aperiodic electron firehose instabilities are thought to be the main instabilities constraining electron temperatures in the solar wind \citep{2008JGRA..113.3103S,2019A&A...627A..76S}.

On the other hand, the solar wind eVDFs can be parameterized  by a superposition of at least two electron components  \citep[e.g.,] [] {1975JGR....80.4181F,1987JGR....92.1075P,2005JGRA..110.9104M,2008JGRA..113.3103S,2017A&A...602A..44L,2019ApJ...870L...6T}, a dense electron component and a tenuous electron component that possess a drift velocity faster than that of the dense component. The latter electron beam can be the halo electron population in the slow solar wind and the strahl electron population in the fast solar wind \citep[e.g.,] [] {1975JGR....80.4181F,2019ApJ...870L...6T}. This electron beam is responsible for the electron heat flux observed in the solar wind \citep[e.g.,] [] {1975JGR....80.4181F}. The electron beam can drive electromagnetic fluctuations through the electron beam-induced instability \citep[or the electron heat flux instability; e.g.,][]{1975JGR....80.4197G,1985JGR....9010815G,1994JGR....9923391G,2017MNRAS.465.1672S,2018MNRAS.480..310S,2019MNRAS.486.4498S,2019ApJ...876..117L,2019ApJ...882L...8L,2019ApJ...870L...6T}, and then these fluctuations can scatter beam electrons, resulting in the nonequilibrium eVDFs toward equilibrium state. Consequently, the electron beam-induced instabilities introduce constraints on the differential drift velocity among different electron populations in the solar wind. Furthermore, the electron beam can destabilize the whistler wave \citep[e.g.,][]{1975JGR....80.4197G}, the electron acoustic wave \citep[e.g.,][]{1984GeoRL..11.1180T,1985JGR....90.6327M,2004PhPl...11.1996S}, and other kinds of plasma waves \citep{1993tspm.book.....G}. The electron beam-driven whistler instability is widely thought of as an effective constraint on the electron beam speed in the solar wind \citep{1977JGR....82.1087G,1994JGR....9923391G}.

Since the electron temperature anisotropy and the electron beam both contribute to the eVDFs in the solar wind, they should be taken into account in the kinetic instability analysis at the same time. The whistler heat flux instability and the electromagnetic electron cyclotron instability are analyzed by combining these two free energy sources in the solar wind \citep{2017MNRAS.466.4928S,2018PhPl...25h2105S,2019MNRAS.483.5642S}; however, these studies only consider the parallel propagation condition. In this study, we investigate both parallel and oblique electron kinetic instabilities driven by the electron temperature anisotropy and the electron beam. Motivated by instabilities in the $(\beta_{e\parallel},~A_e)$ space given in \cite{2008JGRA..113.3103S}, we firstly give comprehensive instability distributions in the same parameter space, where $\beta_{e\parallel}$ is the ratio of the electron parallel thermal pressure to the magnetic pressure. Our results explore the possible electron instability constraints on temperature anisotropies and electron beam velocities in the solar wind. Moreover, we show the presence of a new instability in the regime of $\beta_{e\parallel}\sim0.1-2$ and $A_e<1$, which produces obliquely propagating fast-magnetosonic/whistler waves.

This paper is organized as follows. Section 2 gives the plasma parameters and the theoretical model. Section 3 presents the instability distributions driven by the interplay of the electron temperature anisotropy and the electron beam. The discussion and summary are included in Section 4. In the Appendices, we present the instabilities driven by the electron temperature anisotropy and the electron beam separately.

\section{Plasma parameters and theoretical model} \label{sec:parameter}

In order to show the effects of the electron temperature anisotropy and the electron beam on  excitation of electron kinetic instabilities clearly, we consider a plasma containing three particle components, that is, a proton population with isotropic temperatures, and counter-drifting 
electron populations, i.e., a dense core (subscript “c”) with anisotropic temperatures, and a 
tenuous electron beam (subscript “b”). The proton VDF is assumed to be isotropic nondrifting 
Maxwellian

\begin{equation}
f_p({\bf v}) = \frac{1}{\pi^{3/2}V_{\mathrm{Tp}}^3} \mathrm{exp}\left( -v^2/V_{\mathrm{Tp}}^2\right),
\label{PVDF}
\end{equation}
where $V_{\mathrm{Tp}} \equiv \left(2T_{p}/m_p\right)^{1/2}$ is the proton thermal speed, $m_p$ is the proton mass, and $T_{p}=T_{p\parallel}=T_{p\perp}$ is the proton temperature. Both core and beam electrons are described by drifting bi-Maxwellian VDFs

\begin{widetext}
\begin{eqnarray}
f_{\mathrm{ec}} ({v_\parallel, v_\perp}) =  \frac{1}{\pi^{3/2}V_{\mathrm{Tec\parallel}}V_{\mathrm{Tec\perp}}^2}
 \mathrm{exp}\left[ -\frac{\left(v_\parallel - V_{\mathrm{ec}}\right)^2}{V_{\mathrm{Tec\parallel}}^2} 
 - \frac{v_{\perp}^2}{V_{\mathrm{Tec\perp}}^2} \right],
\end{eqnarray}

and

\begin{eqnarray}
f_{\mathrm{eb}} ({v_\parallel, v_\perp}) =  \frac{1}{\pi^{3/2}V_{\mathrm{Teb\parallel}}V_{\mathrm{Teb\perp}}^2}
 \mathrm{exp}\left[ -\frac{\left(v_\parallel - V_{\mathrm{eb}}\right)^2}{V_{\mathrm{Teb\parallel}}^2} 
 -\frac{v_\perp^2}{V_{\mathrm{Teb\perp}}^2} 
 \right],
\end{eqnarray}
\end{widetext}

respectively. $V_{\mathrm{Tec\parallel}} = \left(2T_{\mathrm{ec\parallel}}/m_e \right)^{1/2}$ and $V_{\mathrm{Tec\perp}} = \left(2T_{\mathrm{ec\perp}}/m_e \right)^{1/2}$ represent the parallel and perpendicular thermal speed of core electrons, respectively. $V_{\mathrm{Teb\parallel}} = \left(2T_{\mathrm{eb\parallel}}/m_e \right)^{1/2}$ and $V_{\mathrm{Teb\perp}} = \left(2T_{\mathrm{eb\perp}}/m_e \right)^{1/2}$ denote the parallel and perpendicular thermal speed of beam electrons, respectively, in a plasma frame fixed to protons. $V_{\mathrm{ec}}$ and $V_{\mathrm{eb}}$ denote the drift velocity of core electrons and beam electrons, respectively. Here we consider the zero net current condition, $n_{\mathrm{ec}}V_{\mathrm{ec}}+n_{\mathrm{eb}}V_{\mathrm{eb}}=0$, and therefore, $V_{\mathrm{ec}}=-n_{\mathrm{eb}}V_{\mathrm{eb}}/n_{\mathrm{ec}}$, where $n_{\mathrm{ec}}$ and $n_{\mathrm{eb}}$ are core and beam electron number densities, respectively.
Note that in this study we consider isotropic beam electrons, $T_{\mathrm{eb}}=T_{\mathrm{eb\parallel}}=T_{\mathrm{eb\perp}}$, and the effects of anisotropic electron beam on electron instabilities will be studied in the future.

In fact, the electron population in the high energy range (normally larger than 100 eV, corresponding to halo and strahl electrons) is usually fitted by the bi-Kappa distribution in the solar wind \citep[e.g.,][]{2010SoPh..267..153P,2016SoPh..291.2165P}. Through statistically analyzing eVDFs in the solar wind, the averaged $\kappa$ index distributes $\sim 5-8$ in the radial distance of 0.35-1.0 AU \citep{2016SoPh..291.2165P}. Since the Kappa distribution can return the Maxwellian distribution as $\kappa\rightarrow\infty$, our plasma model (bi-Maxwellian beam electrons) is the zeroth-order approximation for actual solar wind electrons \citep[neglecting the effects of the suprathermal electrons; also see][]{2017MNRAS.466.4928S,2018PhPl...25h2105S,2019A&A...627A..76S}. Although instability features (e.g., the growth rate, the real frequency, and the unstable wavenumber region) of the electron instability at $\kappa\rightarrow\infty$ quantitatively deviate from the results at small $\kappa$ \citep[e.g., $\kappa=2-4$; see][]{2017A&A...602A..44L,2017MNRAS.464..564L}, the former result can reasonably describe the features of electron instabilities in the solar wind \citep[see][]{2017MNRAS.466.4928S,2018PhPl...25h2105S,2019A&A...627A..76S}. In particular, for the heat flux instability in a Kappa-halo model ($\kappa=8$, $\beta_{\mathrm{ec}}=0.04$, and $\beta_{\mathrm{eb}}=0.36$), both the growth rate and the wave frequency are consistent with the corresponding values in the Maxwellian-halo model \citep{2017MNRAS.465.1672S}. Consequently, our results can give basic properties of electron instability distributions in the solar wind.

According to the solar wind electron observations \citep[e.g,][]{1975JGR....80.4181F,2016SoPh..291.2165P}, we adopt the following magnetic field and plasma parameters: $B_0=5$ nT, $n_p=7.6~\mathrm{cm}^{-3}$, $n_{\mathrm{ec}}=0.965n_p$, $n_{\mathrm{eb}}=0.035n_p$, $T_{p}=T_{\mathrm{ec\parallel}}$, and $T_{\mathrm{eb}}=5T_{\mathrm{ec\parallel}}$.
We consider a typical drift velocity $V_{\mathrm{eb}}=30V_{\mathrm{A}}$ for beam electrons. The drift velocity of core electrons approximates $V_{\mathrm{ec}}\simeq -1.1 V_{\mathrm{A}}$ , where $V_{\mathrm{A}}\simeq 39.6$ km/s. 
Therefore, we can explore electron kinetic instabilities in the $(\beta_{\mathrm{ec\parallel}},A_{\mathrm{ec}})$ space through changing $T_{\mathrm{ec\parallel}}$ and $T_{\mathrm{ec\perp}}$. 

For collective plasma modes in the solar wind, they are described by the kinetic Vlasov equation and Maxwell's equations
\begin{widetext}
\begin{eqnarray}
\partial_t \delta f_s + {\bf v}\cdot\partial_{\bf r} \delta f_s 
+ \frac{q_s}{m_s}\left( {\bf v}\times{\bf B_0} \right)\cdot\partial_{\bf v} \delta f_s
&=&-\frac{q_s}{m_s}\left({\bf E}+{\bf v}\times{\bf B}\right)\cdot\partial_{\bf v} f_s,
\label{Vlasov}
\\
\nabla\times {\bf B} &=& \mu_0{\bf J} + \frac{1}{c^2}\partial_t {\bf E},
\label{B}
\\
\nabla\times {\bf E} &=& -\partial_t {\bf B},
\label{E}
\end{eqnarray}
\end{widetext}

where $f_s$, $q_s$, and $m_s$ represent the distribution function, charge, and mass of the species $s$, respectively. $\delta f_s$ is the perturbed distribution function, ${\bf E}$ is the electric field fluctuation, ${\bf B}$ is the magnetic field fluctuation, and ${\bf B_0}$ is the ambient magnetic field. Under the plane wave assumption, the plasma wave eigenmodes correspond to the solutions of Eqs. (\ref{Vlasov})$-$(\ref{E}). This study will use a newly developed numerical method, PDRK/BO \citep{2016PlST...18...97X,2019ApJ...884...44S,2019CoPhC.244..343X}, to search for all unstable plasma waves in our parameter space. Although PDRK \citep{2016PlST...18...97X} cannot distinguish the wave propagating direction through the input wavevector which is limited as $k_\parallel>0$ and $k_\perp>0$, we can use the output wave frequency $\omega_r$ to identify the wave direction, i.e., $\omega_r>0$ for wave propagating along ${\bf B_0}$, and $\omega_r<0$ for wave propagating against ${\bf B_0}$. Besides the wave frequency and the growth rate, this study will also explore the electromagnetic responses of unstable waves. For example, we use the absolute value of $B_y/B_x$ and the argument defined by $\left(\omega_r/|\omega_r|\right)\left(B_y/B_x\right)$ to characterize the wave polarization.

\section{Instabilities driven by interplay of electron temperature anisotropy and electron beam}  \label{sec:results}

\begin{figure*}[htbp]
\centering
\epsscale{1.0}
\plotone{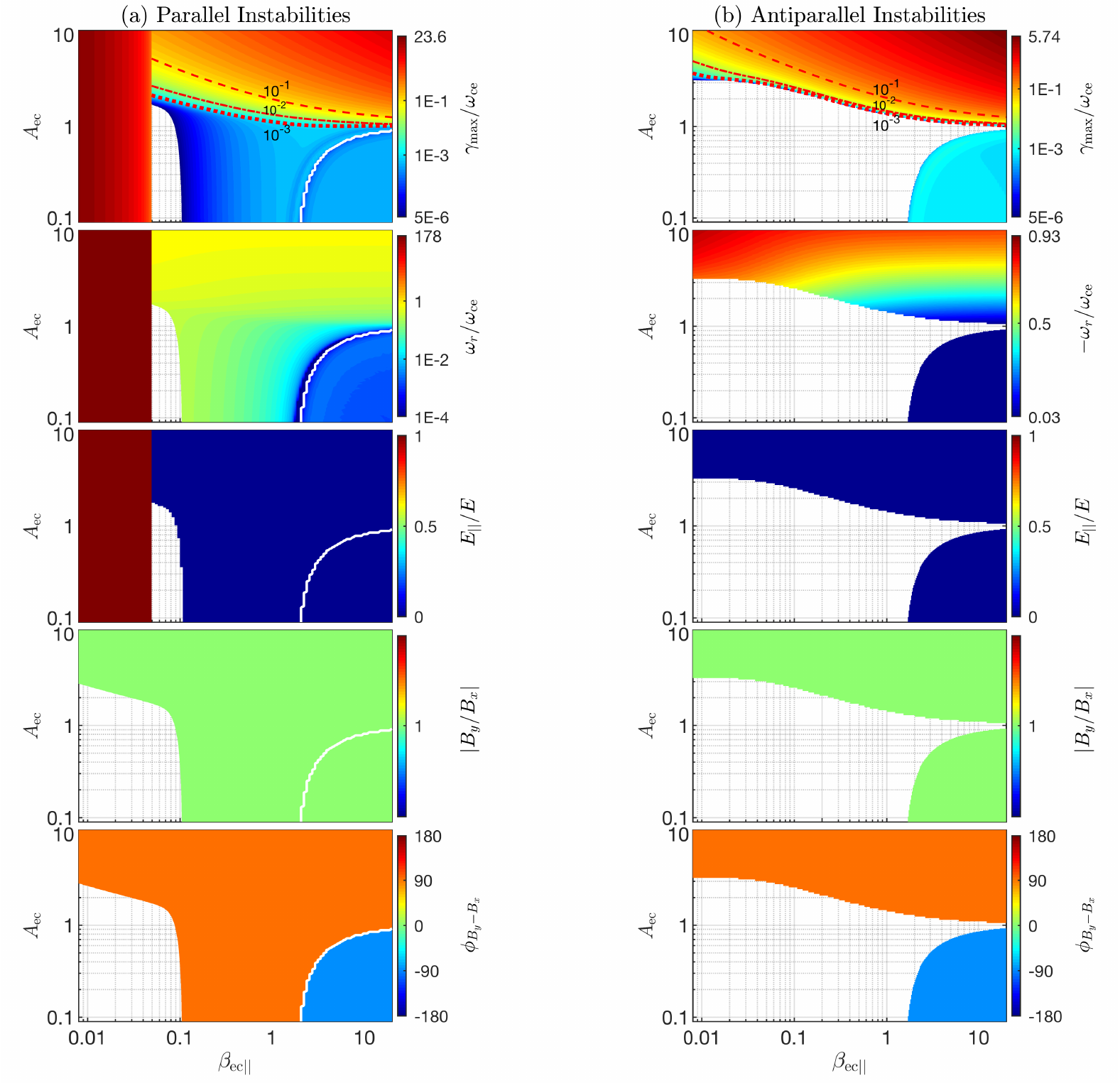}
\caption{The ($\beta_{\mathrm{ec\parallel}},A_{\mathrm{ec}}$) distributions of (a) parallel and (b) antiparallel electron instabilities induced by the electron temperature anisotropy and the electron beam with $V_{\mathrm{eb}}=30V_{\mathrm{A}}$. Top pair of panels (maximum growth rate, $\gamma_{\mathrm{max}}/\omega_{\mathrm{ce}}$), second pair of panels (the real frequency, $\omega_r/\omega_{\mathrm{ce}}$), third pair of panels (the ratio of parallel to total electric field, $E_\parallel/E$), fourth pair of panels (the absolute value of $B_y/B_x$, $|B_y/B_x|$), and bottom pair of panels (the phase difference between $B_x$ and $B_y$, $\phi_{B_y-B_x}\equiv (\omega_r/|\omega_r|)\times \mathrm{arg}\left(B_y/B_x\right)$). The dotted, dashed-dotted, and dashed lines in the top pair of panels represent the threshold value of $A_{\mathrm{ec}}$ corresponding to $\gamma_{\mathrm{max}}/\omega_{\mathrm{ce}}=10^{-3}$, $10^{-2}$, and $10^{-1}$, respectively. The solid lines in the left panels denote boundaries between the electron beam-driven whistler instability and the periodic electron firehose instability.
\label{fig:fig1}}
\end{figure*}

\begin{figure*}[htbp]
\centering
\epsscale{1.0}
\plotone{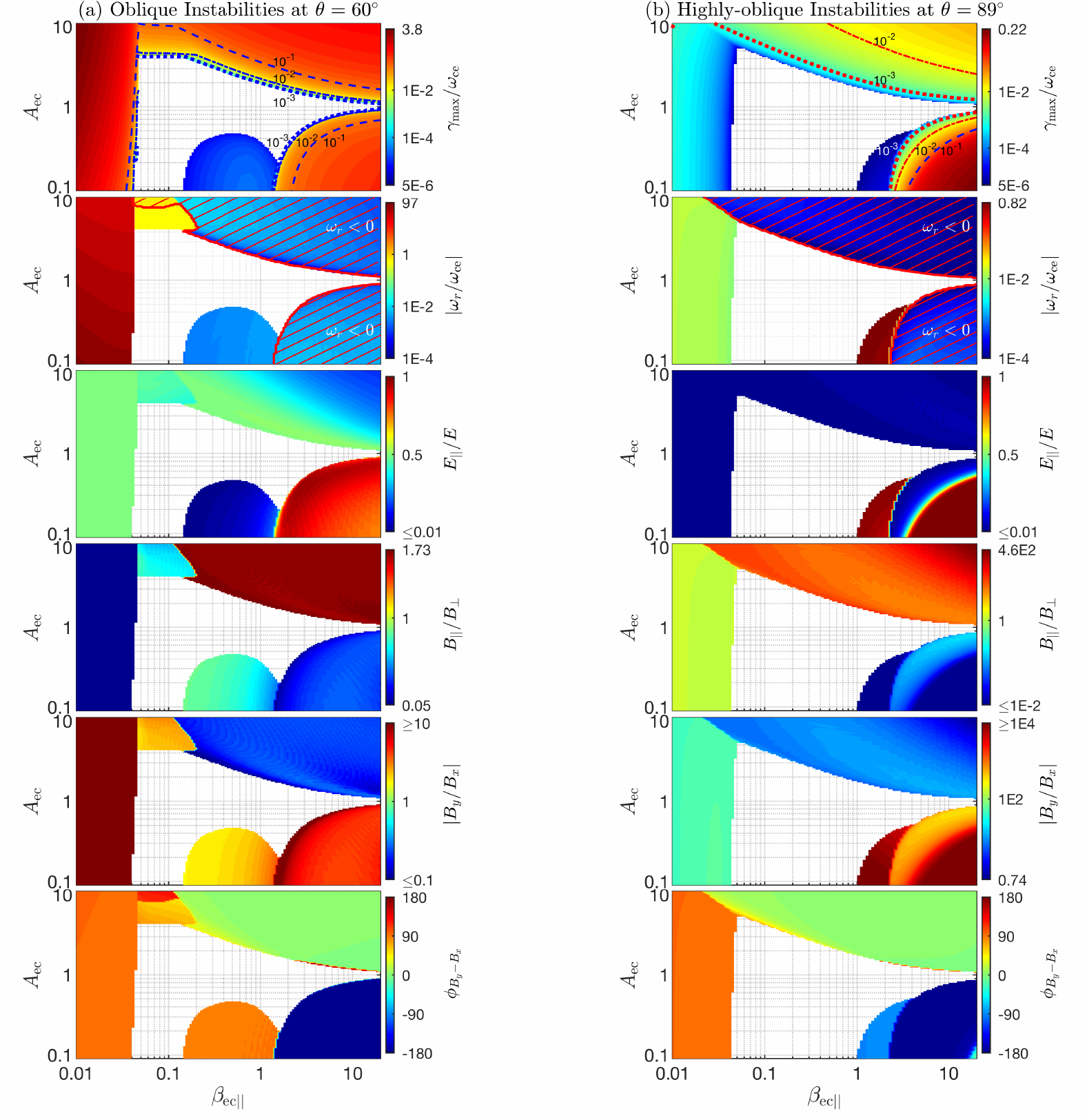}
\caption{The ($\beta_{\mathrm{ec\parallel}},A_{\mathrm{ec}}$) distributions of electron instabilities induced by the electron temperature anisotropy and the electron beam with $V_{\mathrm{eb}}=30V_{\mathrm{A}}$: (a) $\theta=60^\circ$, and (b) $\theta=89^\circ$. 
Top pair of panels ($\gamma_{\mathrm{max}}/\omega_{\mathrm{ce}}$), second pair of panels ($|\omega_r/\omega_{\mathrm{ce}}|$), third pair of panels ($E_\parallel/E$), fourth pair of panels ($B_\parallel/B_\perp$), fifth pair of panels ($|B_y/B_x|$), and bottom pair of panels ($\phi_{B_y-B_x}\equiv (\omega_r/|\omega_r|)\times \mathrm{arg}\left(B_y/B_x\right)$). 
The dotted, dashed-dotted, and dashed lines in the top pair of panels represent the threshold value of $A_{\mathrm{ec}}$ corresponding to $\gamma_{\mathrm{max}}/\omega_{\mathrm{ce}}=10^{-3}$, $10^{-2}$, and $10^{-1}$, respectively. 
\label{fig:fig2}}
\end{figure*}

Figures \ref{fig:fig1} and \ref{fig:fig2}  show the $(\beta_{\mathrm{ec\parallel}},A_{\mathrm{ec}})$ distributions for the strongest instability driven by the interplay of the electron temperature anisotropy and the electron beam. For parallel and antiparallel propagation shown in Figure \ref{fig:fig1}, there are four kinds of instabilities, namely, the electron beam-driven electron acoustic instability in the low-$\beta_{\mathrm{ec\parallel}}$ regime, the electron beam-driven whistler instability, the electromagnetic electron cyclotron instability, and the periodic electron firehose instability. For oblique propagation ($\theta=60^\circ$ and $89^\circ$) shown in Figure \ref{fig:fig2}, six kinds of instabilities exist: the electron beam-driven electron magnetoacoustic and whistler instabilities in the low-$\beta_{\mathrm{ec\parallel}}$ regime, the electron mirror instability, the aperiodic electron firehose instability, the oblique fast-magnetosonic/whistler instability ($\theta=60^\circ$), and the ordinary-mode instability ($\theta=89^\circ$). These instabilities dominate different parameter regimes.

As $\beta_{\mathrm{ec\parallel}}\lesssim0.05$, the dominant instability corresponds to the electron beam-driven electron acoustic instability in the parallel direction, the electron beam-driven electron magnetoacoustic instability at $\theta=60^\circ$, and the electron beam-driven whistler instability at $\theta=89^\circ$. In the parallel direction (Figure \ref{fig:fig1}a), the electron beam triggers electrostatic ($E_\parallel=E$) electron acoustic waves with maximum growth rate $\gamma_{\mathrm{max}}\sim24\omega_{\mathrm{ce}}$ and wave frequency $\omega_r\sim 178\omega_{\mathrm{ce}}$. At $\theta=60^\circ$ (Figure \ref{fig:fig2}a), the electron beam drives electron magnetoacoustic waves having $\gamma_{\mathrm{max}}\sim3.8\omega_{\mathrm{ce}}$, $\omega_r\sim97\omega_{\mathrm{ce}}$, $E_\parallel\simeq0.5E$, $B_\parallel\ll B_\perp$, and $|B_y|\gg| B_x|$. At $\theta=89^\circ$, the electron beam mainly excites whistler waves that have $\gamma_{\mathrm{max}}\sim10^{-4}\omega_{\mathrm{ce}}$, $\omega_r\sim0.01\omega_{\mathrm{ce}}$, $E_\parallel\ll E$, $B_\parallel\sim B_\perp$, and $|B_y|\gg |B_x|$.

As $\beta_{\mathrm{ec\parallel}}\gtrsim0.05$, the electron beam-driven whistler instability nearly dominates the regime with $A_{\mathrm{ec}}$ between the solid and dotted lines (Figure \ref{fig:fig1}a), where the excited whistler waves are propagating along $\bm{B}_0$. The solid line represents the boundary between the electron beam-driven whistler instability and the periodic electron firehose instability. The dotted line can approximately distinguish the electron beam-driven whistler instability from the electromagnetic electron cyclotron instability. However, due to both the electron beam-driven whistler instability and the electromagnetic electron cyclotron instability resulting in the same whistler mode wave, it is hard to strictly distinguish the boundary between these two instabilities. It should be noted that their boundary is more clear in the quasi-linear theory \citep{2020MNRAS.492.3529S}.

In the regime where $\beta_{\mathrm{ec\parallel}}\gtrsim0.05$ and $A_{\mathrm{ec}}>1$, the electromagnetic electron cyclotron instability arises in parallel and antiparallel directions (Figure \ref{fig:fig1}), and the electron mirror instability appears in oblique directions ($\theta=60^\circ$ and $89^\circ$; Figure \ref{fig:fig2}). The electromagnetic electron cyclotron instability generates right-hand polarized whistler waves ($\phi_{B_y-B_x}=90^\circ$). If there is no electron beam, parallel and antiparallel electromagnetic electron cyclotron instabilities have the same growth rates. Due to the electron beam along $\bm{B}_0$, three $A_{\mathrm{ec}}$ threshold values (corresponding to $\gamma_{\mathrm{max}}/\omega_{\mathrm{ce}}=10^{-3},~10^{-2}$ and $10^{-1}$, indicated by the dashed, dashed-dotted and dotted lines in Figure \ref{fig:fig1}) in the parallel electromagnetic electron cyclotron instability are lower than that in the antiparallel instability. The unstable regime of the parallel electromagnetic electron cyclotron instability is broader than the unstable regime for corresponding antiparallel instability. Moreover, mirror-mode waves have nonzero frequency $\omega_r<0$. Due to mirror-mode waves generated by anisotropic core electrons, their nonzero frequencies come from the Doppler shifting frequency $\omega_r\sim V_{\mathrm{ec}}k<0$ resulting from backstreaming core electrons.

In the regime of $\beta_{\mathrm{ec\parallel}}\gtrsim2$ and $A_{\mathrm{ec}}<1$, parallel and antiparallel periodic electron firehose instabilities dominate in parallel and antiparallel directions, and both instabilities produce left-hand polarized waves ($\phi_{B_y-B_x}=-90^\circ$; see Figure \ref{fig:fig1}). Moreover, the parallel instability has the growth rate $\sim 10^{-4}\omega_{\mathrm{ce}}$ much smaller than that in the antiparallel instability ($\sim10^{-3}\omega_{\mathrm{ce}}$). The aperiodic electron firehose instability dominates in oblique directions (see Figures 2 and 3).

In the regime of $\beta_{\mathrm{ec\parallel}}\sim0.2-2$ and $A_{\mathrm{ec}}<1$, an oblique fast-magnetosonic/whistler instability arises at $\theta=60^\circ$ (Figure \ref{fig:fig2}a). It generates oblique fast-magnetosonic/whistler waves having $\omega_r\sim10^{-2}\omega_{\mathrm{ce}}$, $E_\parallel=10^{-3}E$, $B_\parallel\sim B_\perp$, $|B_y|>|B_x|$, and $\phi_{B_y-B_x}\simeq 90^\circ$.

In a regime where $\beta_{\mathrm{ec\parallel}}\sim1-2$ and $A_{\mathrm{ec}}<1$, the ordinary-mode instability dominates at $\theta=89^\circ$ (Figure \ref{fig:fig2}b). The excited ordinary-mode waves have $\omega_r\sim0.8\omega_{\mathrm{ce}}$, $E_\parallel\simeq E$, $|B_y|\gg |B_x|$, and  $\phi_{B_y-B_x}\simeq-90^\circ$.

Furthermore, Figures \ref{fig:fig3}$-$\ref{fig:fig8} exhibit dependence of aforementioned instabilities on the $A_{\mathrm{ec}}$ and/or $\theta$. Here we focus on instabilities at $\beta_{\mathrm{ec\parallel}}\gtrsim 0.05$.

\subsection{Electron beam-driven whistler instability}

\begin{figure}
\plotone{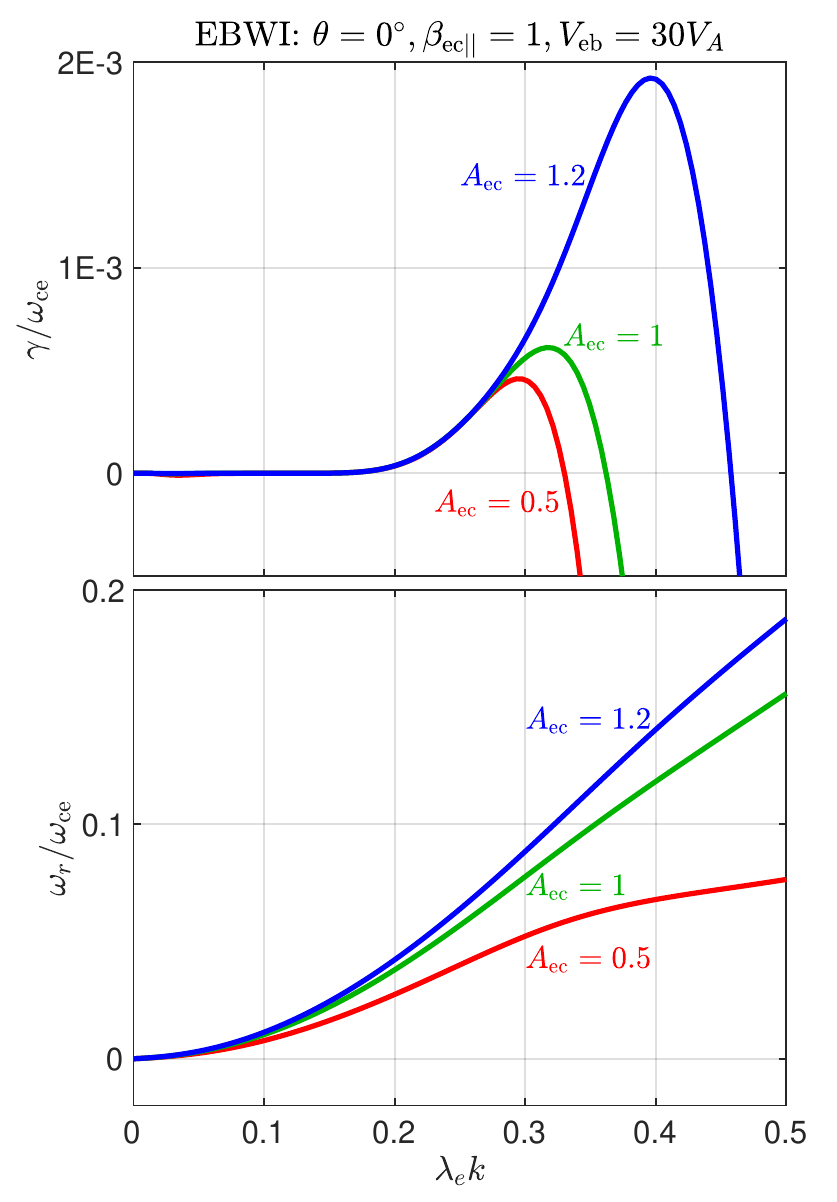}
\caption{EBWI in a plasma where $\beta_{\mathrm{ec\parallel}}=1$ and $V_{\mathrm{eb}}=30V_{\mathrm{A}}$. Top panel (the growth rate, $\gamma/\omega_{\mathrm{ce}}$), and bottom panel (the real frequency, $\omega_r/\omega_{\mathrm{ce}}$).
The red, green, and blue lines represent $A_{\mathrm{ec}}=0.5$, $A_{\mathrm{ec}}=1$, and $A_{\mathrm{ec}}=1.2$, respectively. EBWI=Electron Beam-driven Whistler Instability.
\label{fig:fig3}}
\end{figure}

Figure \ref{fig:fig3} shows dependence of the electron beam-driven whistler instability on the $A_{\mathrm{ec}}$. This instability enhances as $A_{\mathrm{ec}}$ increases \citep[see also][]{2018PhPl...25h2105S}, for example, $\gamma/\omega_{\mathrm{ce}}\simeq0.4\times10^{-3}$ at $A_{\mathrm{ec}}=0.5$, $\gamma/\omega_{\mathrm{ce}}\simeq0.6\times10^{-3}$ at $A_{\mathrm{ec}}=1$, $\gamma/\omega_{\mathrm{ce}}\simeq1.9\times10^{-3}$ at $A_{\mathrm{ec}}=1.2$. Also, the wave frequency and wavenumber at the position of the maximum growth rate become larger with increasing $A_{\mathrm{ec}}$, for example, $\omega_r/\omega_{\mathrm{ce}}\simeq0.05$ and $\lambda_ek\simeq0.29$ at $A_{\mathrm{ec}}=0.5$, $\omega_r/\omega_{\mathrm{ce}}\simeq0.08$ and $\lambda_ek\simeq0.31$ at $A_{\mathrm{ec}}=1$, and $\omega_r/\omega_{\mathrm{ce}}\simeq0.14$ and $\lambda_ek\simeq0.4$ at $A_{\mathrm{ec}}=1.2$.

\subsection{Electromagnetic electron cyclotron instability and electron mirror instability}

\begin{figure*}[htbp]
\centering
\epsscale{1.0}
\plotone{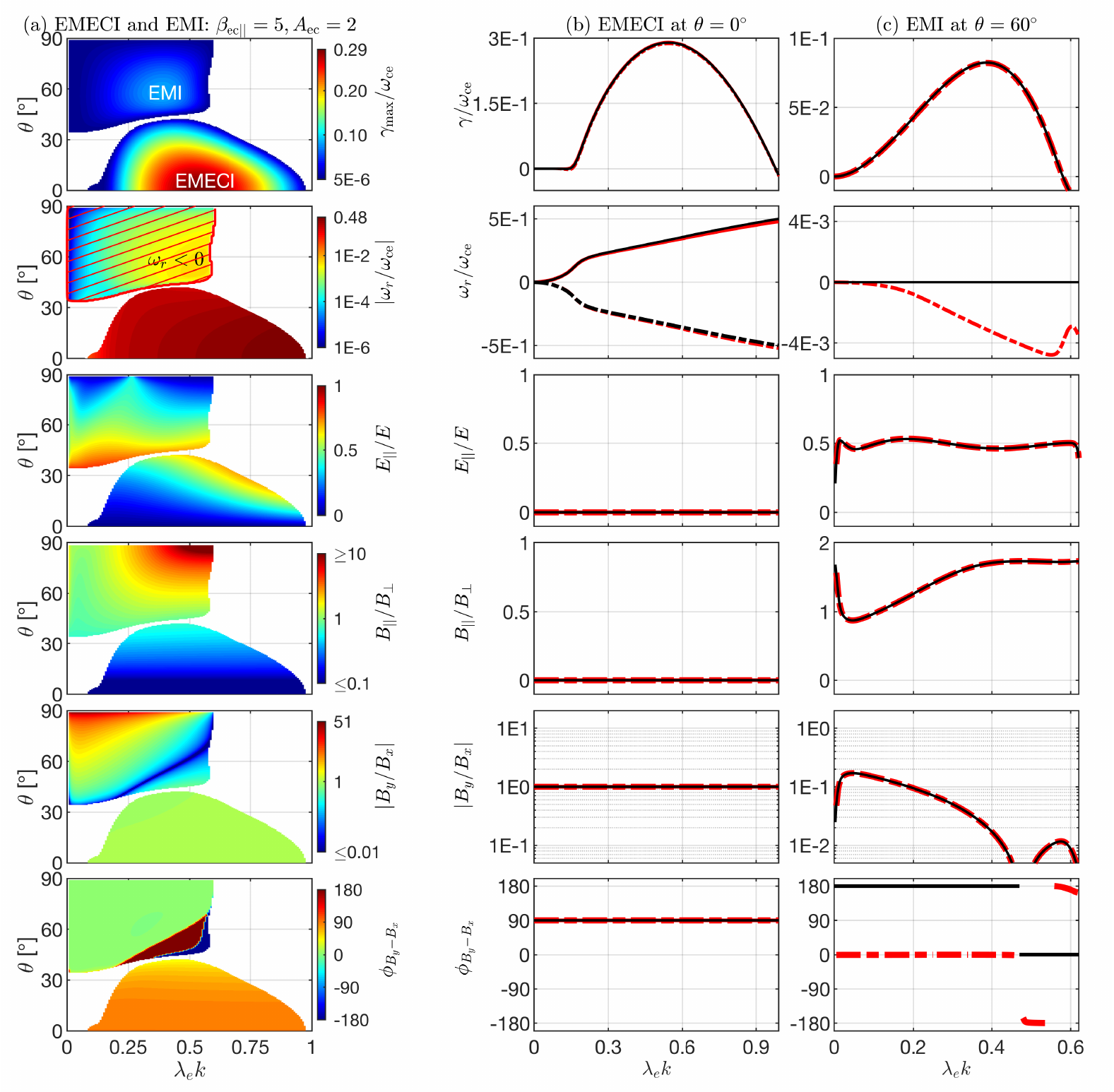}
\caption{
Electron perpendicular temperature anisotropy instabilities in a plasma where $\beta_{\mathrm{ec\parallel}}=5$ and $A_{\mathrm{ec}}=2$: (a) EMECI and EMI in the ($k,\theta$) space; (b) EMECI at $\theta=0^\circ$; and (c) EMI at $\theta=60^\circ$.
Top set of panels ($\gamma_{\mathrm{max}}/\omega_{\mathrm{ce}}$ or $\gamma/\omega_{\mathrm{ce}}$), second set of panels ($|\omega_r/\omega_{\mathrm{ce}}|$ or $\omega_r/\omega_{\mathrm{ce}}$), third set of panels ($E_\parallel/E$), fourth set of panels ($B_\parallel/B_\perp$), fifth set of panels ($|B_y/B_x|$), and bottom set of panels ($\phi_{B_y-B_x}\equiv (\omega_r/|\omega_r|)\times \mathrm{arg}\left(B_y/B_x\right)$). 
The red and black lines denote the unstable waves with and without streaming electrons. The solid and dashed-dotted lines represent forward and backward unstable waves, respectively.
EMECI=ElectroMagnetic Electron Cyclotron Instability; EMI=Electron Mirror Instability.
\label{fig:fig4}}
\end{figure*}

Figure \ref{fig:fig4} presents the distributions of the $A_{\mathrm{ec}}>1$ instability in a plasma where $\beta_{\mathrm{ec\parallel}}=5$ and $A_{\mathrm{ec}}=2$. Figure \ref{fig:fig4}(a) shows that the parallel electromagnetic electron cyclotron instability dominates in the region of $\theta\sim 0^\circ-35^\circ$, and the electron mirror instability distributes at angles $\sim 35^\circ-90^\circ$. For the electromagnetic electron cyclotron instability, it excites forward whistler waves having $\omega_r\sim0.48\omega_{\mathrm{ce}}$, $E_\parallel/E\sim0-0.7$, $B_\parallel/B_\perp\sim0-0.6$, $|B_y|\simeq|B_x|$, and $\phi_{B_y,B_x}\simeq90^\circ$. The electron mirror instability generates backward electron mirror-mode waves with $\omega_r/\omega_{\mathrm{ce}}\sim-10^{-3}$, $E_\parallel/E\sim0.2-0.8$, $B_\parallel/B_\perp\gtrsim1$, $|B_y|/|B_x|\sim0.01-51$, and $\phi_{B_y-B_x}\simeq0^\circ$ ($\phi_{B_y-B_x}\simeq\pm180^\circ$ at $\theta\sim35^\circ-65^\circ$ and $\lambda_ek\sim0.5$).

The strongest electromagnetic electron cyclotron and electron mirror instabilities are shown in Figures \ref{fig:fig4}(b) and \ref{fig:fig4}(c), which also give these two kinds of instabilities in motionless plasmas. Figure \ref{fig:fig4}(b) shows that the parallel electromagnetic electron cyclotron instability is slightly stronger than the antiparallel instability. Moreover, due to the Doppler shifting frequency induced by backstreaming core electrons, the wave frequency of parallel whistler waves is smaller than that of antiparallel waves. Figure \ref{fig:fig4}(c) shows that except $\omega_r$ and $\phi_{B_y-B_x}$, the electron mirror instabilities in two plasma environments (the plasma with electron beams and the motionless plasma) have the same distributions of the growth rate and electromagnetic relations. Note that $\phi_{B_y-B_x}\equiv \mathrm{arg}(B_y/B_x)$ is used for mirror-mode waves in motionless plasma. Besides, comparing maximum growth rates in Figures \ref{fig:fig4}(b) and \ref{fig:fig4}(c), $\gamma\simeq0.29\omega_{\mathrm{ce}}$ of the electromagnetic electron cyclotron instability is nearly four times larger than $\gamma\simeq0.08\omega_{\mathrm{ce}}$ of the electron mirror instability.

\subsection{Periodic and aperiodic electron firehose instability}

Figure \ref{fig:fig5} presents the distributions of the electron firehose instability in a plasma where $\beta_{\mathrm{ec\parallel}}=5$ and $A_{\mathrm{ec}}=0.6$. From the $(k,\theta)$ distributions in Figure \ref{fig:fig5}(a), we see that the periodic electron firehose instability dominates in the region of $\theta\lesssim10^\circ$, where $E_\parallel/E\lesssim0.1$, $|B_y|/|B_x|\sim1$, and $\phi_{B_y-B_x}\simeq-90^\circ$, and the aperiodic electron firehose instability dominates in a broad angle region ($\theta\gtrsim10^\circ$), where $E_\parallel/E\gtrsim0.5$, $|B_y|/|B_x|\gtrsim1$, and $\phi_{B_y-B_x}\simeq0^\circ/-180^\circ$.

Figure \ref{fig:fig5}(b) exhibits the periodic electron firehose instability at $\theta=0^\circ$. Comparing the growth rate ($\gamma_{\mathrm{max}}\simeq3\times10^{-4}\omega_{\mathrm{ce}}$) in motionless plasmas, the antiparallel periodic electron firehose instability ($\gamma_{\mathrm{max}}\simeq 10^{-3}\omega_{\mathrm{ce}}$) is greatly enhanced, and the parallel instability ($\gamma_{\mathrm{max}}\simeq 10^{-4}\omega_{\mathrm{ce}}$) is reduced. For the aperiodic electron firehose instability shown in Figure \ref{fig:fig5}(c), its growth rate is weakly affected by the effects of the electron beam. However, the wave frequency is considerably affected by both forward flowing beam electrons and backstreaming core electrons, and consequently, with increasing $\lambda_ek$, the wave frequency changes from $\omega_r<0$ to $\omega_r>0$ and eventually $\omega_r<0$. If there are no streaming electrons, the aperiodic electron firehose instability excites zero-frequency waves.

\subsection{Oblique Fast-magnetosonic/Whistler instability}

Figure \ref{fig:fig6} presents the distributions of an oblique fast-magnetosonic/whistler instability in a plasma with $\beta_{\mathrm{ec\parallel}}=0.2$ and $A_{\mathrm{ec}}=0.2$. Figure \ref{fig:fig6}a shows that this instability arises in a wide angle region $\theta\sim 15^\circ-89^\circ$, and the growth rate approaches maximum as $\theta\simeq85^\circ$. With increasing $\lambda_ek$, the wave frequency increases from $\omega\sim0.1\omega_{\mathrm{cp}}$ to $\omega\sim10\omega_{\mathrm{cp}}$. There also exist intermittent stable regions where the wave frequency is nearly $n\omega_{\mathrm{cp}}$ ($n$ denotes the natural number), and the appearance of these stable regions is due to the strong cyclotron resonance damping occurring at $\omega=n\omega_{\mathrm{cp}}$. The electromagnetic features of these unstable fast-magnetosonic/whistler waves are $E_\parallel/E\lesssim0.03$, $|B_y|/|B_x|\sim0.1-8$, and $\phi_{B_y-B_x}\simeq90^\circ$.

Figure \ref{fig:fig6}(b) exhibits the oblique fast-magnetosonic/whistler instability at $\theta=85^\circ$. It clearly shows that unstable waves nearly locate between $n\omega_{\mathrm{cp}}$ and $(n+1)\omega_{\mathrm{cp}}$, and the strongest instability appears at $\omega_r\simeq6.5\omega_{\mathrm{cp}}$. To further identify the wave mode, Figure \ref{fig:fig6}(b) also gives the dispersion relation of the fast-magnetosonic/whistler mode wave in the plasma fluid model \citep{2015PhPl...22d2115Z,2019PhPl...26b2108H}, and these two dispersion relations are nearly the same. 

\subsection{Ordinary-mode instability}

Figure \ref{fig:fig7} presents the distributions of the ordinary-mode instability in a plasma with $\beta_{\mathrm{ec\parallel}}=2$ and $A_{\mathrm{ec}}=0.1$. The $(k,\theta)$ distributions in Figure \ref{fig:fig7}a show that the instability is limited in the region of $\theta\sim 88.3^\circ-89.7^\circ$, where the unstable waves have $|\omega_r|/\omega_{\mathrm{ce}}\sim0.62-0.82$, $E_\parallel/E\sim0.85-0.99$, $|B_y|/|B_x|\gg1$, and $\phi_{B_y-B_x}\simeq-90^\circ$. From Figure \ref{fig:fig7}(b), we see that the effects of the electron beam can induce imbalanced growth rates between forward and backward ordinary-mode waves. Comparing the growth rate $\gamma\simeq10^{-5}\omega_{\mathrm{ce}}$ in motionless plasmas, $\gamma\simeq1.5\times10^{-5}\omega_{\mathrm{ce}}$ of forward waves is larger than $\gamma\simeq0.5\times10^{-5}\omega_{\mathrm{ce}}$ of backward waves. Also, the electron beam results in $E_\parallel/E$ decreasing (increasing) and $|B_y|/|B_x|$ increasing (decreasing) for forward (backward) waves. However, the electron beam weakly affects the distributions of $\omega_r$ and $\phi_{B_y-B_x}$.

\begin{figure*}[htbp]
\centering
\epsscale{1.0}
\plotone{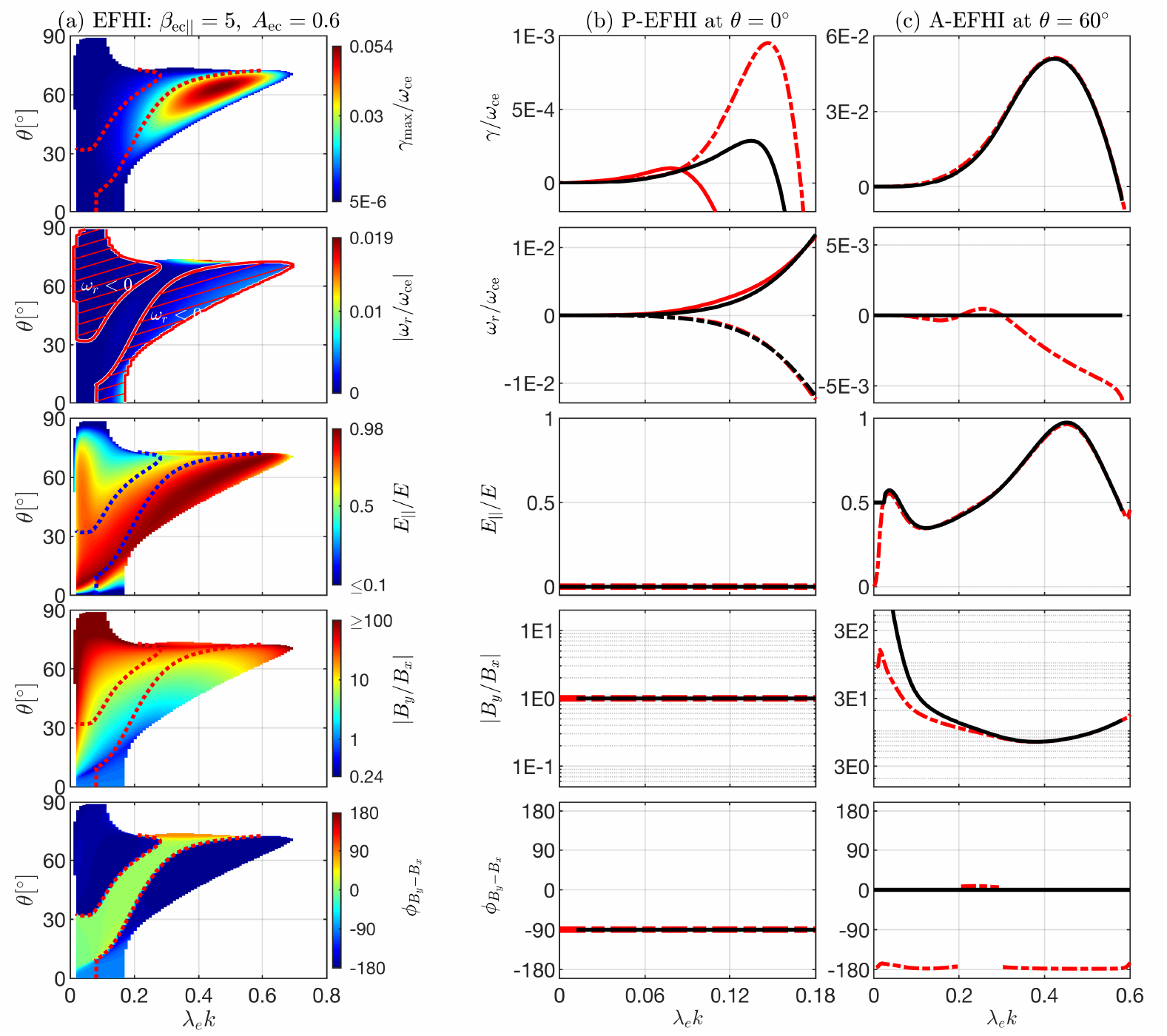}
\caption{
Electron parallel temperature anisotropy instabilities in a plasma where $\beta_{\mathrm{ec\parallel}}=5$ and $A_{\mathrm{ec}}=0.6$: 
(a) EFHI in the ($k,\theta$) space; (b) P-EFHI at $\theta=0^\circ$; and (c) A-EFHI at $\theta=60^\circ$.
Top set of panels ($\gamma_{\mathrm{max}}/\omega_{\mathrm{ce}}$ or $\gamma/\omega_{\mathrm{ce}}$), second set of panels ($|\omega_r/\omega_{\mathrm{ce}}|$ or $\omega_r/\omega_{\mathrm{ce}}$), third set of panels ($E_\parallel/E$), fourth set of panels ($|B_y/B_x|$), and bottom set of panels ($\phi_{B_y-B_x}\equiv (\omega_r/|\omega_r|)\times \mathrm{arg}\left(B_y/B_x\right)$). 
The red and black lines denote the unstable waves with and without streaming electrons. The solid and dashed-dotted lines represent forward and backward unstable waves, respectively.
EFHI=Electron FireHose Instability; A-EFHI=Aperiodic Electron FireHose Instability; P-EFHI=Periodic Electron FireHose Instability.
\label{fig:fig5}}
\end{figure*}

\begin{figure*}[htbp]
\centering
\epsscale{1.0}
\plotone{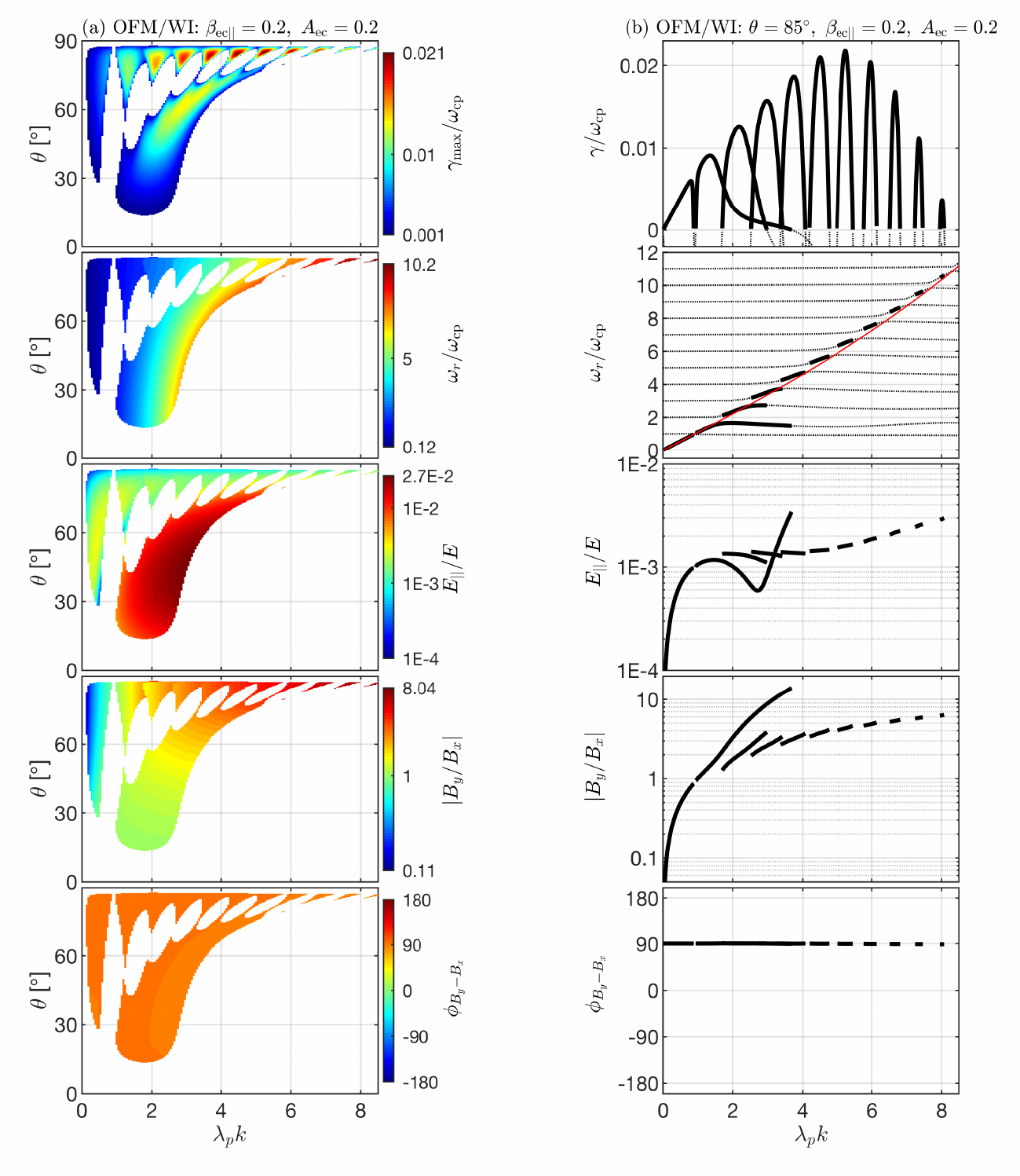}
\caption{OFM/WI in a plasma where $\beta_{\mathrm{ec\parallel}}=0.2$ and $A_{\mathrm{ec}}=0.2$: the instability distributions in the ($k,\theta$) space, and (b) the instability at $\theta=85^\circ$.
Top pair of panels ($\gamma_{\mathrm{max}}/\omega_{\mathrm{cp}}$ or $\gamma/\omega_{\mathrm{cp}}$), second pair of panels ($|\omega_r/\omega_{\mathrm{cp}}|$ or $\omega_r/\omega_{\mathrm{cp}}$), third pair of panels ($E_\parallel/E$), fourth pair of panels ($|B_y/B_x|$), and bottom pair of panels ($\phi_{B_y-B_x}\equiv (\omega_r/|\omega_r|)\times \mathrm{arg}\left(B_y/B_x\right)$). 
The red line denotes the fast-magnetosonic/whistler wave in the fluid model. 
OFM/WI=Oblique Fast-Magnetosonic/Whistler Instability.
\label{fig:fig6}}
\end{figure*}

\begin{figure*}[htbp]
\centering
\epsscale{0.98}
\plotone{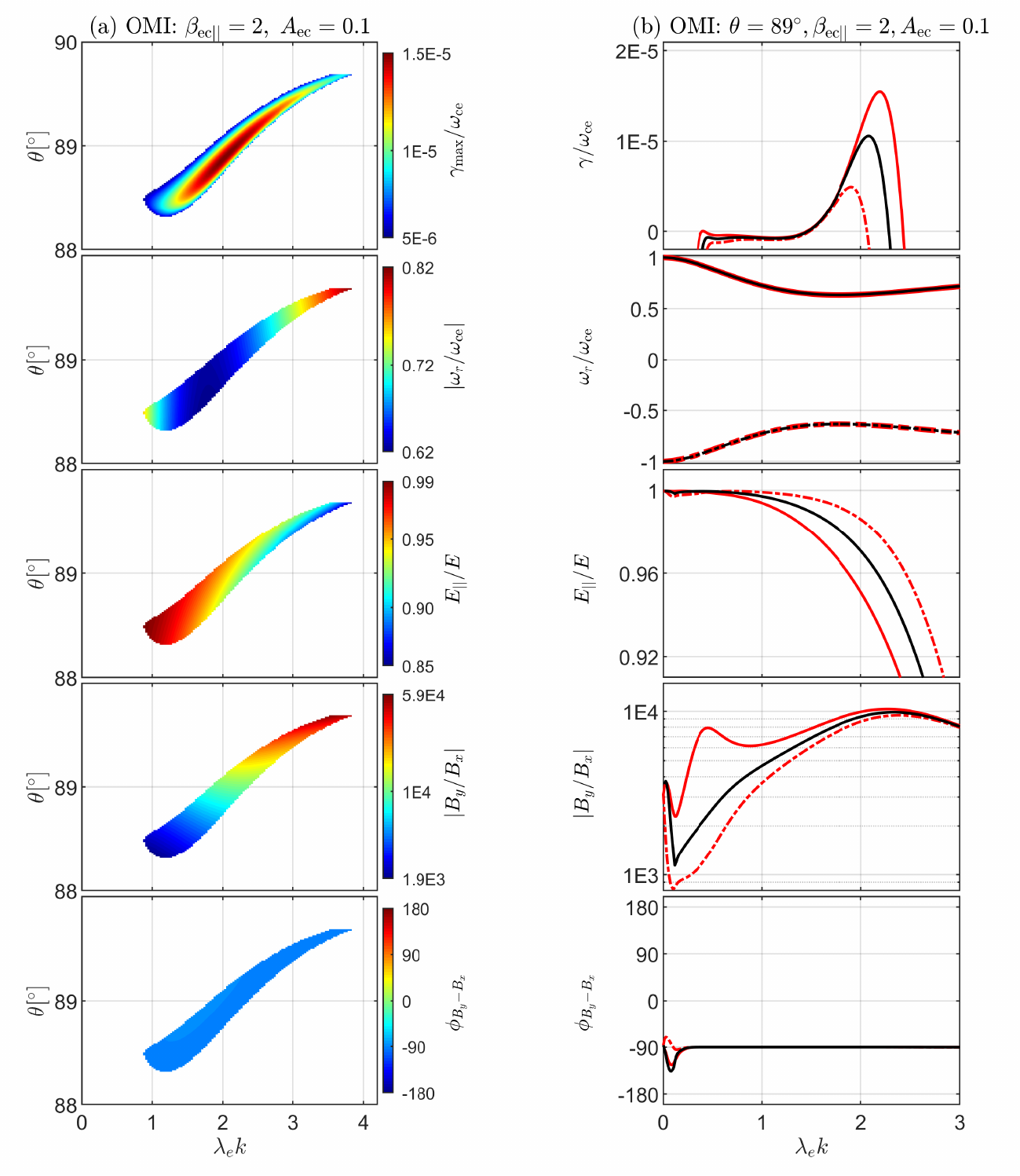}
\caption{OMI in a plasma where $\beta_{\mathrm{ec\parallel}}=2$ and $A_{\mathrm{ec}}=0.1$: 
(a) the instability distributions in the ($k,\theta$) space; and (b) the instability at $\theta=89^\circ$.
Top pair of panels ($\gamma_{\mathrm{max}}/\omega_{\mathrm{ce}}$ or $\gamma/\omega_{\mathrm{ce}}$), second pair of panels ($|\omega_r/\omega_{\mathrm{ce}}|$ or $\omega_r/\omega_{\mathrm{ce}}$), third pair of panels ($E_\parallel/E$), fourth pair of panels ($|B_y/B_x|$), and bottom pair of panels ($\phi_{B_y-B_x}\equiv (\omega_r/|\omega_r|)\times \mathrm{arg}\left(B_y/B_x\right)$). 
The red and black lines denote the unstable waves with and without streaming electrons. The solid and dashed-dotted lines represent forward and backward unstable waves, respectively.
OMI=Ordinary-Mode Instability.
\label{fig:fig7}}
\end{figure*}

\section{Discussions and summary}

Both electron temperature anisotropy and electron beam contribute to the nonequilibrium eVDFs in the solar wind. These nonequilibrium velocity distributions will be unstable to drive different kinds of electron instabilities. In turn, these instabilities can shape actual velocity distribution in the solar wind. Previous studies have shown that the electron temperature distributions are mainly constrained by the electromagnetic electron cyclotron instability and the aperiodic electron firehose instability \citep{1996JGR...10110749G,2003PhPl...10.3571G,2008JGRA..113.3103S}, and the differential flow among different electron populations is constrained by the electron beam-induced whistler instability \citep{1977JGR....82.1087G,1994JGR....9923391G,2019A&A...627A..76S}.

In this study, we investigate electron kinetic instabilities induced by both the electron temperature anisotropy and the electron beam in the same parameter space. Moreover, to complement the instability distribution, we present electron instabilities resulting from the electron temperature anisotropy and the electron beam separately in the Appendices. Therefore, our results can give a comprehensive overview for the instability constraint on the solar wind electron dynamics. In particular, we find that the $\beta_{e\parallel}$ is an important parameter to determine which type of electron instability is triggered.

\subsection{Constraint on the electron beam}

The electron beam mainly results in three kinds of instabilities dominating in different $\beta_{\mathrm{ec\parallel}}$ regimes, i.e., the electron beam-driven electron acoustic/magnetoacoustic instability in low-$\beta_{\mathrm{ec\parallel}}$ ($\lesssim 0.05$) regime, the electron beam-driven whistler instability in $\beta_{\mathrm{ec\parallel}}\gtrsim0.05$ regime, and the oblique fast-magnetosonic/whistler instability in the regime of $\beta_{\mathrm{ec\parallel}}\sim0.2-2$ and $A_{\mathrm{ec}}<1$.

In the $\beta_{\mathrm{ec\parallel}}\lesssim 0.05$ regime, the electron beam-driven acoustic/magnetoacoustic instability dominates. Furthermore, from the $(\beta_{\mathrm{ec\parallel}},V_{\mathrm{eb}})$ distribution of the electron beam instability (Figure \ref{fig:figA3}), we see that the threshold value of $V_{\mathrm{ebThre}}$ approximates as $V_{\mathrm{ebThre}}\simeq 3V_{\mathrm{Tec}}$. The electron beam-driven acoustic/magnetoacoustic instability is very strong ($\gamma>\omega_{\mathrm{ce}}$), and it can effectively limit the electron beam velocity nearby the electron thermal velocity. Moreover, the electron beam drives oblique whistler waves (see Figure \ref{fig:figA4}) in the low-$\beta_{\mathrm{ec\parallel}}$ regime; however, this instability ($\gamma\sim0.01\omega_{\mathrm{ce}}$) is much weaker than the electron beam-driven acoustic/magnetoacoustic instability. 

In the $\beta_{\mathrm{ec\parallel}}\gtrsim 0.05$ regime, the electron beam can directly excite parallel whistler waves (see Figures \ref{fig:fig1} and \ref{fig:figA3}$-$\ref{fig:figA4}). This whistler instability is induced by the normal cyclotron resonance between whistler waves and antipropagating electrons \citep[e.g.,][]{2019ApJ...886..136V}. When $\beta_{\mathrm{ec\parallel}}$ increases from $\sim0.1$ to $\sim6$, the threshold value of  $V_{\mathrm{eb}}$ can decrease from $\sim24V_{\mathrm{A}}$ to $\sim6V_{\mathrm{A}}$ (Figure \ref{fig:figA3}). Also, we find the growth rate in the electron beam-driven whistler instability is increasing with $A_{\mathrm{ec}}$, which is consistent with previous results given by \cite{2018MNRAS.480..310S}.

In the regime of $\beta_{\mathrm{ec\parallel}} \sim0.1-2$ and $A_{\mathrm{ec}}<1$, in addition to the electron beam-driven whistler instability, the electron beam can excite an oblique fast-magnetosonic/whistler instability (Figures \ref{fig:fig2}(a) and \ref{fig:fig6}). To compare these two instabilities, Figure 8 presents their $(V_{\mathrm{eb}}, \theta)$ distributions. It shows that the electron beam-driven whistler instability dominates at small angles ($\theta\lesssim 5^\circ$), and the oblique fast-magnetosonic/whistler instability controls at large angles  ($\theta\gtrsim 15^\circ$). Moreover, the former instability arises in the $14V_{\mathrm{A}}\lesssim V_{\mathrm{eb}}\lesssim 50V_{\mathrm{A}}$ range, and disappears as  $V_{\mathrm{eb}}\gtrsim 50V_{\mathrm{A}}$. The latter instability appears at a lower electron beam speed ($V_{\mathrm{eb}}\simeq 7V_{\mathrm{A}}$), and its growth rate increases with $V_{\mathrm{eb}}$. Therefore, the oblique fast-magnetosonic/whistler instability can play an important role in constraining the electron beam in the $A_{\mathrm{ec}}<1$ solar wind plasmas.

\begin{figure}
\plotone{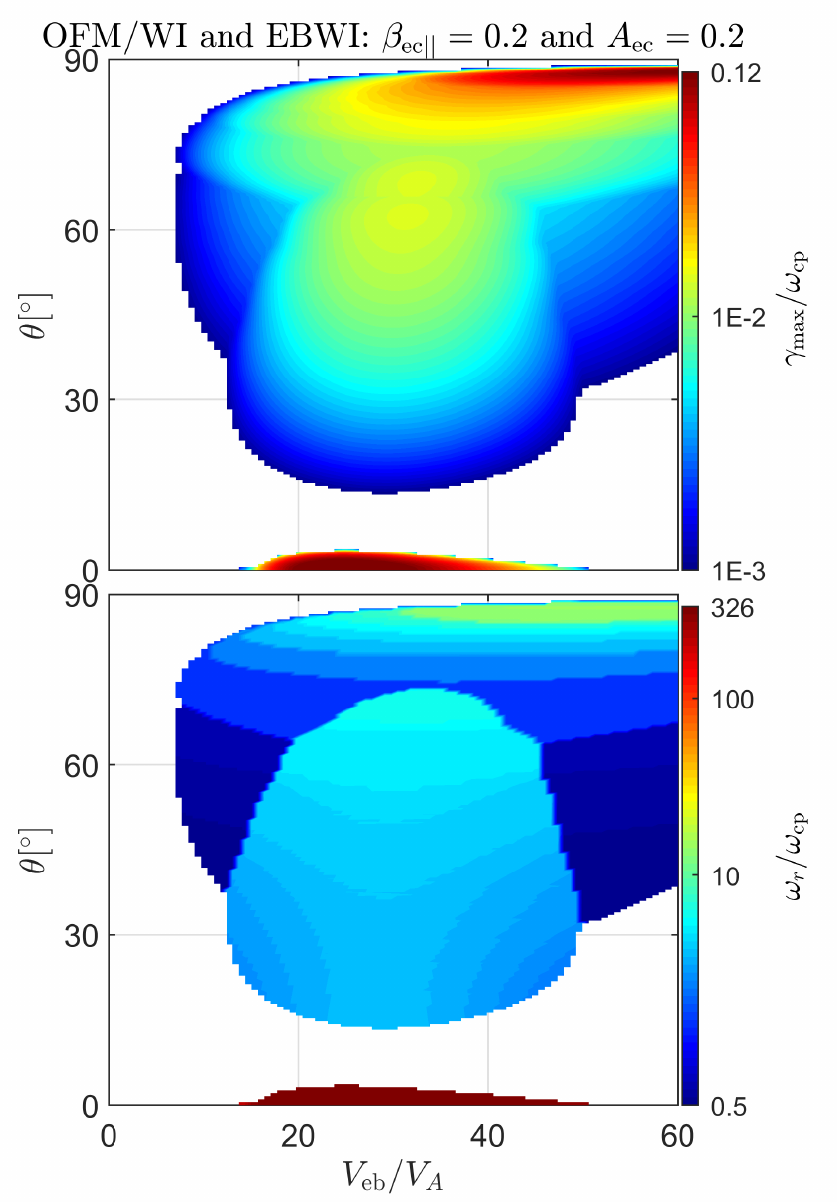}
\caption{The ($V_{\mathrm{eb}}$, $\theta$) distributions of EBWI and OFM/WI in a plasma where $\beta_{\mathrm{ec\parallel}}=0.2$ and $A_{	\mathrm{ec}}=0.2$. 
Top panel ($\gamma_{\mathrm{max}}/\omega_{\mathrm{cp}}$), and bottom panel ($\omega_r/\omega_{\mathrm{cp}}$).
EBWI=Electron Beam-driven Whistler Instability; OFM/WI=Oblique Fast-Magnetosonic/Whistler Instability.
\label{fig:fig8}}
\end{figure}

Recently, \cite{2019ApJ...886..136V} considered both core and beam electron distributions with isotropic temperatures in the solar wind, and found that oblique fast-magnetosonic/whistler waves can be excited by the electron beam with $V_{\mathrm{eb}}\gtrsim 3V_{\mathrm{Tec}}$ as $\beta_{\mathrm{ec}}<0.1$ and by the electron beam with  $V_{\mathrm{eb}}\gtrsim 2-3V_{\mathrm{Ae}}\sim86-129V_{\mathrm{A}}$ as $\beta_{\mathrm{ec}}\simeq 0.1-2$, where $V_{\mathrm{Ae}}$ is the electron Alfv\'en speed. This oblique fast-magnetosonic/whistler instability is different from our oblique fast-magnetosonic/whistler instability in the regime of $\beta_{\mathrm{ec\parallel}} \sim0.1-2$ and $A_{\mathrm{ec}}<1$. Both the electron beam and the electron temperature anisotropy provide free energies to excite our instability, and the threshold velocity of the electron beam, $V_{\mathrm{eb}}\sim7V_{\mathrm{A}}$, is much smaller than the electron thermal speed. Note that the electron beam-driven whistler instability in the $\beta_{\mathrm{ec\parallel}}\lesssim 0.1$ regime can trigger oblique whistler waves, and the instability threshold is $V_{\mathrm{eb}}\gtrsim 3V_{\mathrm{Tec}}$, in accordance with the mechanism proposed by \cite{2019ApJ...886..136V}.

\begin{figure*}[htbp]
\centering
\epsscale{1.0}
\plotone{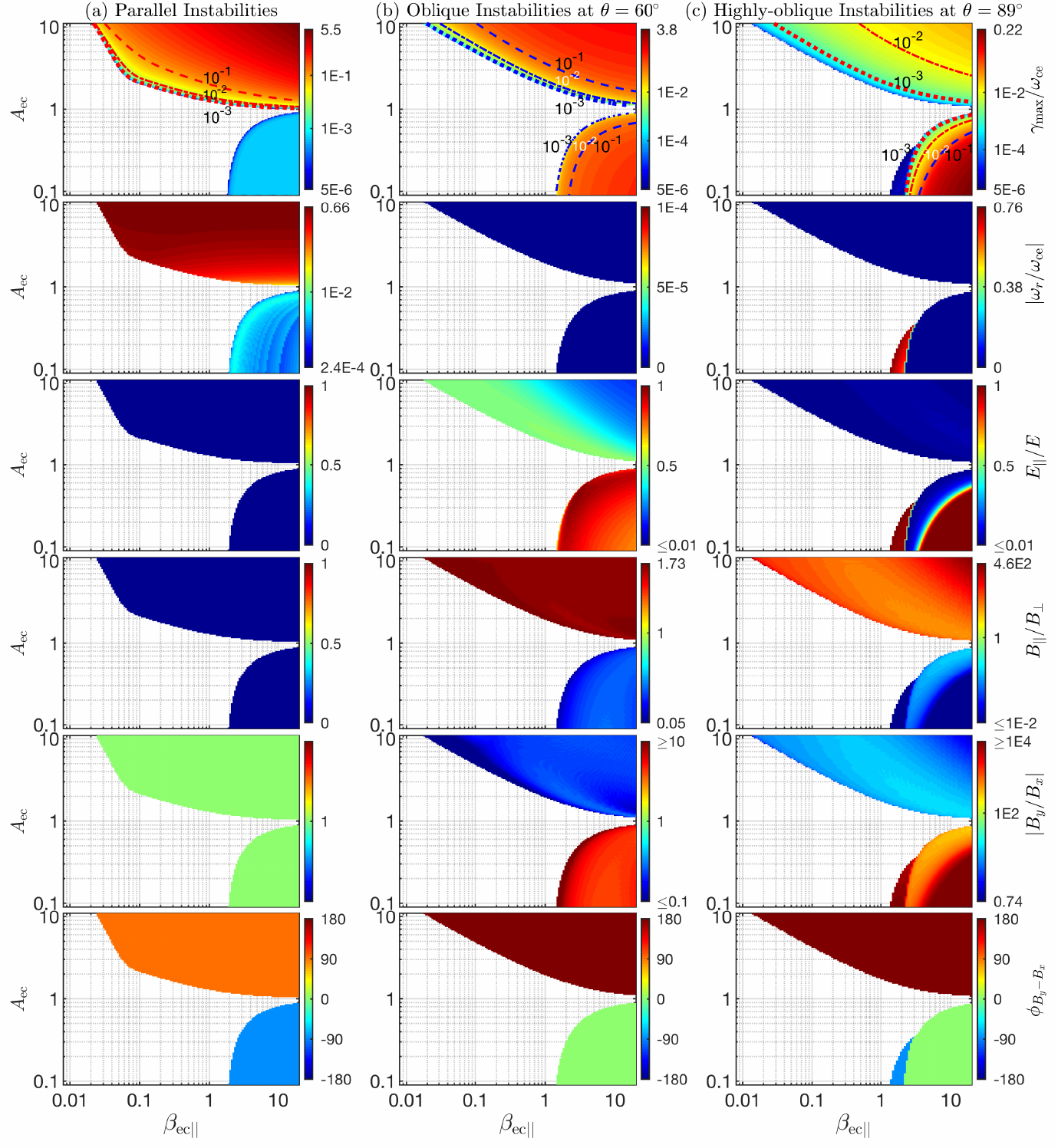}
\caption{The ($\beta_{\mathrm{ec\parallel}},A_{\mathrm{ec}}$) distributions of electron instabilities induced by the electron temperature anisotropy: (a) $\theta=0^\circ$; (b) $\theta=60^\circ$; and (c) $\theta=89^\circ$.
Top set of panels ($\gamma_{\mathrm{max}}/\omega_{\mathrm{ce}}$), second set of panels ($|\omega_r/\omega_{\mathrm{ce}}|$), third set of panels ($E_\parallel/E$), fourth set of panels ($B_\parallel/B_\perp$), fifth set of panels ($|B_y/B_x|$), and bottom set of panels ($\phi_{B_y-B_x}\equiv (\omega_r/|\omega_r|)\times \mathrm{arg}\left(B_y/B_x\right)$, and $\phi_{B_y-B_x}\equiv \mathrm{arg}\left(B_y/B_x\right)$ for zero-frequency waves). 
The dotted, dashed-dotted, and dashed lines in top set of panels represent the threshold value of $A_{\mathrm{ec}}$ corresponding to $\gamma_{\mathrm{max}}/\omega_{\mathrm{ce}}=10^{-3}$, $10^{-2}$, and $10^{-1}$, respectively. 
\label{fig:figA1}}
\end{figure*}

\begin{figure*}[htbp]
\centering
\epsscale{1.0}
\plotone{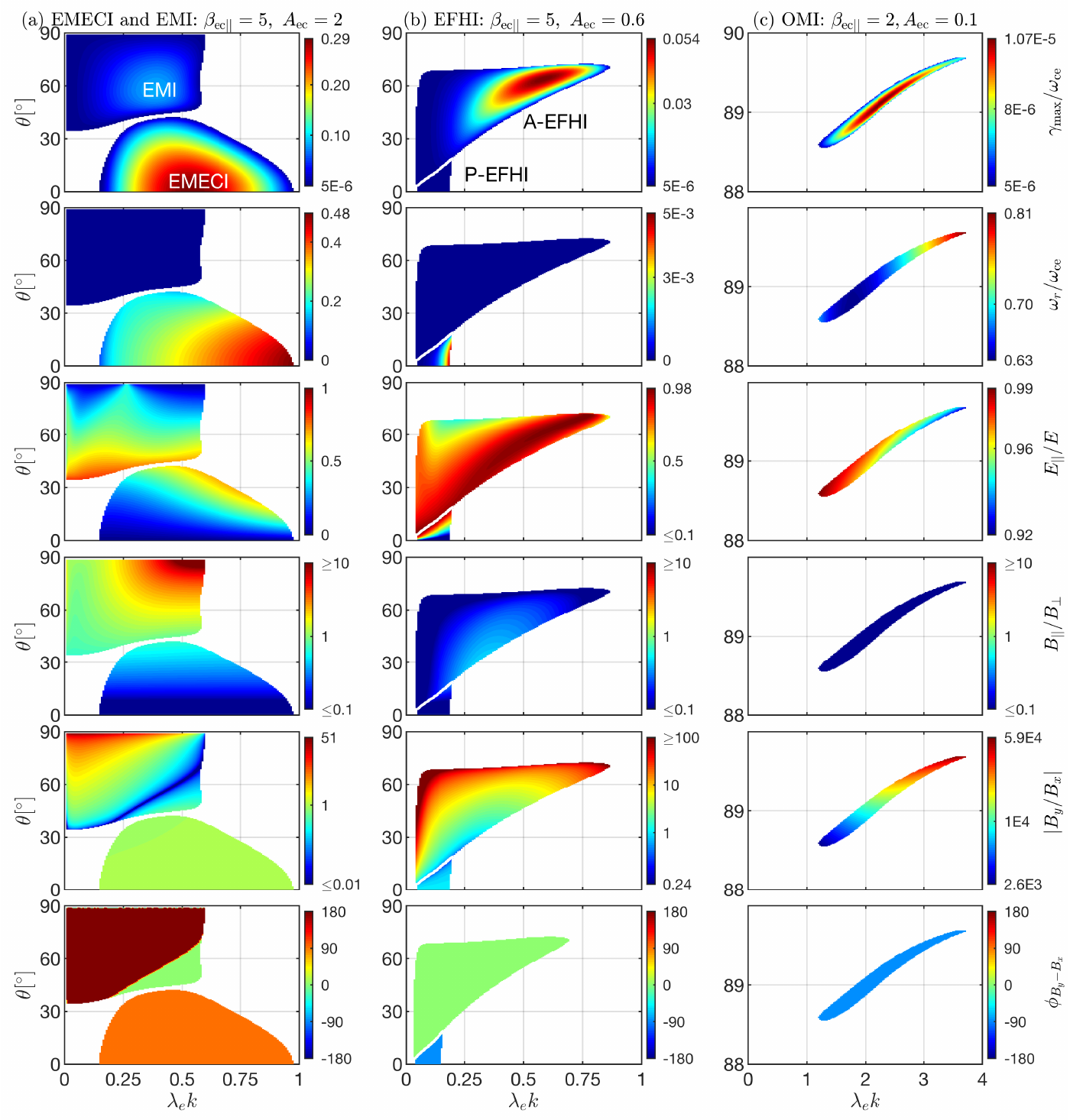}
\caption{
The ($k,\theta$) distributions for (a) EMECI and EMI at $\beta_{\mathrm{ec\parallel}}=5$ and $A_{\mathrm{ec}}=2$, (b) EFHI at $\beta_{\mathrm{ec\parallel}}=5$ and $A_{\mathrm{ec}}=0.6$, and (c) OMI at $\beta_{\mathrm{ec\parallel}}=2$ and $A_{\mathrm{ec}}=0.1$.
Top set of panels ($\gamma_{\mathrm{max}}/\omega_{\mathrm{ce}}$), second set of panels ($|\omega_r/\omega_{\mathrm{ce}}|$), third set of panels ($E_\parallel/E$), fourth set of panels ($B_\parallel/B_\perp$), fifth set of panels ($|B_y/B_x|$), and bottom set of panels ($\phi_{B_y-B_x}\equiv (\omega_r/|\omega_r|)\times \mathrm{arg}\left(B_y/B_x\right)$, and $\phi_{B_y-B_x}\equiv \mathrm{arg}\left(B_y/B_x\right)$ for EMI and A-EFHI). 
EFHI=Electron FireHose Instability; EMECI=ElectroMagnetic Electron Cyclotron Instability; EMI=Electron Mirror Instability; OMI=Ordinary-Mode Instability.
\label{fig:figA2}}
\end{figure*}

\begin{figure*}[htbp]
\centering
\epsscale{1.0}
\plotone{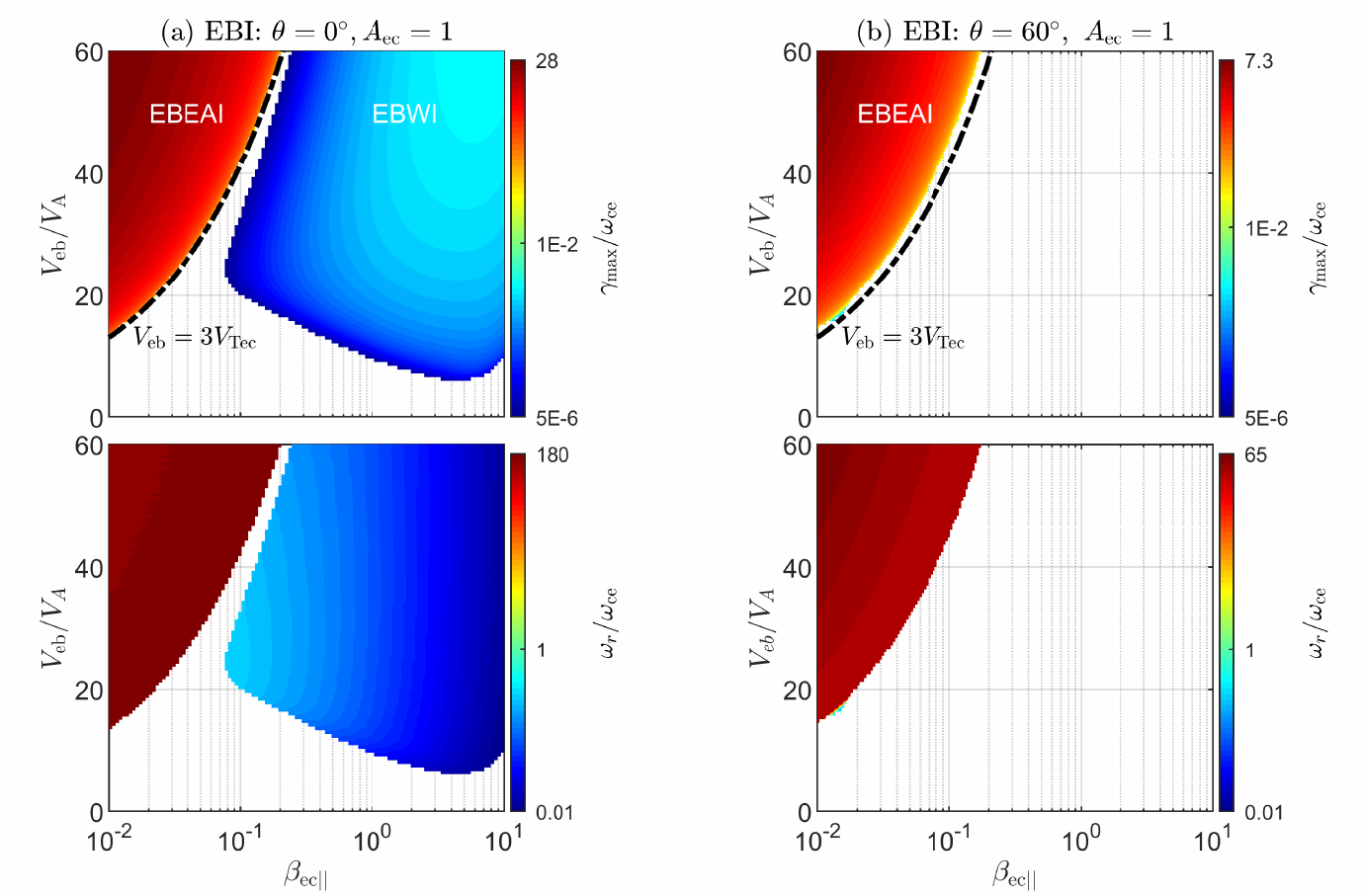}
\caption{
The ($k,\theta$) distributions of electron instabilities induced by the electron beam at (a) $\theta=0^\circ$ and (b) $\theta=60^\circ$, where $A_{\mathrm{ec}}=1$. 
Top pair of panels ($\gamma_{\mathrm{max}}/\omega_{\mathrm{ce}}$), and bottom pair of panels ($\omega_r/\omega_{\mathrm{ce}}$).
EBI=Electron Beam-driven Instability; EBEAI=Electron Beam-driven Electron Acoustic/magnetoacoustic Instability; EBWI=Electron beam-driven Whistler Instability. The black lines denote $V_{\mathrm{ebThre}}= 3V_{\mathrm{Tec}}$. 
\label{fig:figA3}}
\end{figure*}

\subsection{Constraint on the electron temperature anisotropy}

In the $\beta_{\mathrm{ec\parallel}}\gtrsim 0.05$ and $A_{\mathrm{ec}}>1$ regime, there are two kinds of instabilities: the electromagnetic electron cyclotron instability and the electron mirror instability. The growth rate in the electromagnetic electron cyclotron instability is nearly three times larger than that in the electron mirror instability. Hence, the electromagnetic electron cyclotron instability is a dominant constraint on the electron perpendicular temperature anisotropy \citep{2008JGRA..113.3103S}. 

In the $\beta_{\mathrm{ec\parallel}}\gtrsim 0.05$ and $A_{\mathrm{ec}}<1$ regime, the parallel electron temperature anisotropy can induce the periodic electron firehose instability, the aperiodic electron firehose instability, and the ordinary-mode instability. The aperiodic-type electron firehose instability is much stronger than the periodic-type instability, which is consistent with previous results \citep{2000JGR...10527377L,2003PhPl...10.3571G}. However, both periodic and aperiodic electron firehose instabilities disappear in the regime of $\beta_{\mathrm{ec\parallel}}\sim1-2$, where the ordinary-mode instability arises. Therefore, the aperiodic electron firehose and ordinary-mode instabilities are responsible for constraining the electron parallel temperature anisotropy.

Moreover, in a motionless plasma, the electron temperature anisotropy produces symmetric growth rates for forward and backward unstable waves excited by the electromagnetic electron cyclotron instability, the periodic electron firehose instability, or the ordinary-mode instability. The growth rates of counter-propagating waves become asymmetric in the presence of the electron beam. The effects of the electron beam can enhance (reduce) the forward (backward) electromagnetic electron cyclotron instability, the backward (forward) periodic electron firehose instability, and the backward (forward) ordinary-mode instability. As the electron beam speed increases, these asymmetric distributions are more evident. Also, due to Doppler frequency shift induced by streaming electrons, both the electron mirror instability and the periodic electron firehose instability excite the waves with nonzero frequency. 

\begin{figure*}[htbp]
\centering
\epsscale{0.98}
\plotone{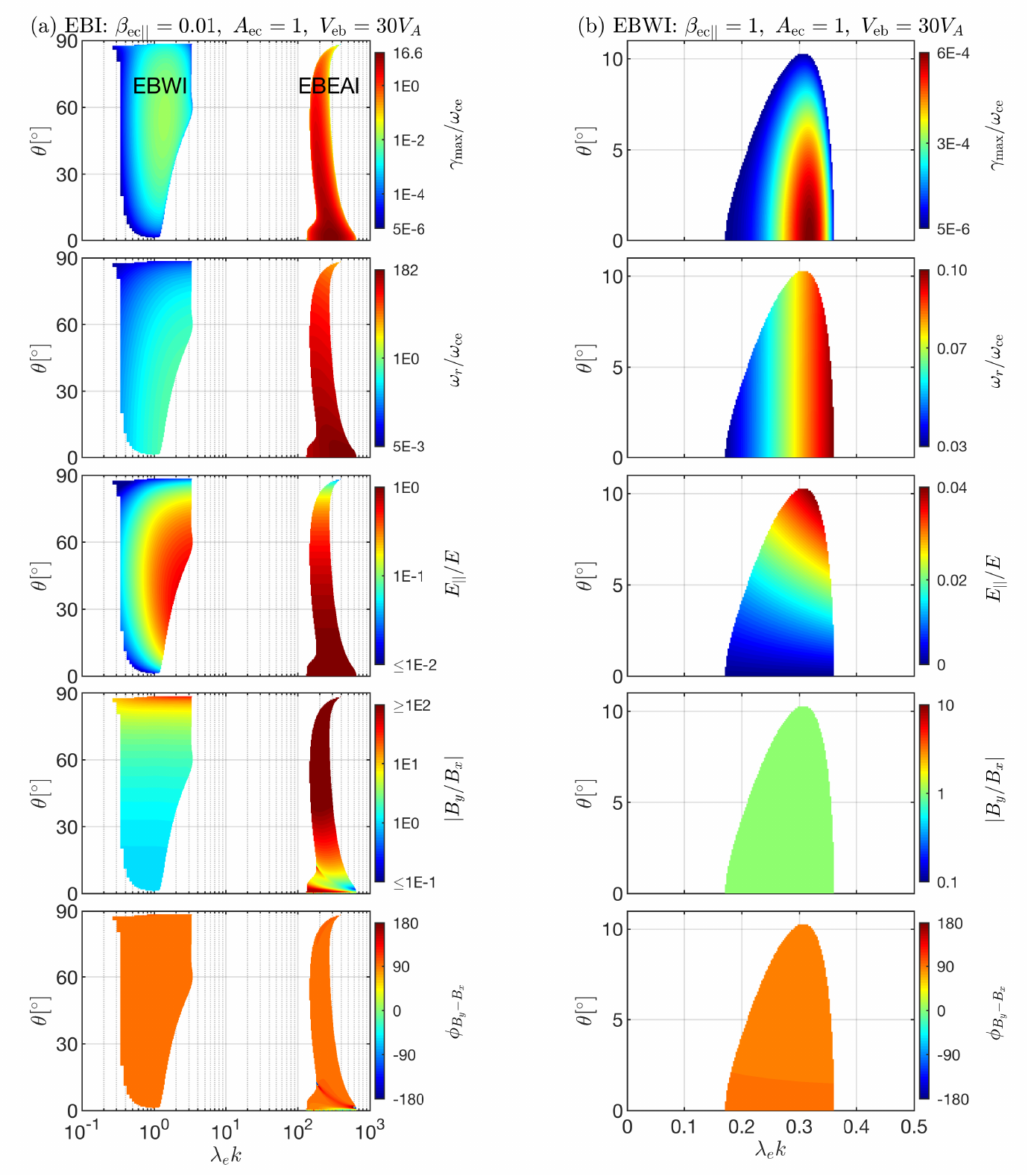}
\caption{
The ($k,\theta$) distributions of (a) EBWI and EBEAI at $\beta_{\mathrm{ec\parallel}}=0.01$ and $V_{\mathrm{eb}}=30V_{\mathrm{A}}$ and (b) EBWI at $\beta_{\mathrm{ec\parallel}}=1$ and $V_{\mathrm{eb}}=30V_{\mathrm{A}}$. 
Top pair of panels ($\gamma_{\mathrm{max}}/\omega_{\mathrm{ce}}$), second pair of panels ($|\omega_r/\omega_{\mathrm{ce}}|$), third pair of panels ($E_\parallel/E$), fourth pair of panels ($|B_y/B_x|$), and bottom pair of panels ($\phi_{B_y-B_x}\equiv (\omega_r/|\omega_r|)\times \mathrm{arg}\left(B_y/B_x\right)$). 
EBI=Electron Beam-driven Instability; EBEAI=Electron Beam-driven Electron Acoustic/magnetoacoustic Instability; EBWI=Electron beam-driven Whistler Instability.
\label{fig:figA4}}
\end{figure*}

In this study, we merely consider temperature anisotropy of core electrons. Actually, both the core and beam electron populations have temperature anisotropy \citep[e.g.,][]{2015A&A...582A.124L,2017MNRAS.466.4928S,2018MNRAS.480..310S,2019ApJ...871..237S}. \cite{2018MNRAS.480..310S} have explored the effects of the core and beam electron anisotropy on the electromagnetic electron cyclotron instability and the heat flux instability. They found that for the electromagnetic electron cyclotron instability triggered by anisotropic core electrons, its growth rate is increasing with the electron beam speed, however, the growth rate decreases with increasing electron beam speed in the instability driven by anisotropic beam electrons. For the whistler heat flux instability, the instability strength is enhanced as $A_{\mathrm{ec}}$ increases (also see Figure 3 in this study), but it is reduced as $A_{\mathrm{eb}}$ increases \citep{2018MNRAS.480..310S}. Therefore, the electron instability is strongly dependent on both the core and beam electron temperature anisotropy. We will present a comprehensive investigation for the electron instability under varying $A_{\mathrm{eb}}$ and $A_{\mathrm{ec}}$.

\subsection{Linear versus quasi-linear theory predictions}

When the electron instability induced by the electron temperature anisotropy and/or the electron beam can amplify electromagnetic waves, the energy continuously redistributes between the waves and particles, and the electron velocity distribution can evolve to a quasi-stable state. This dynamical process can be divided into a linear growing stage and a nonlinear saturation stage \citep[e.g.,][]{2017RvMPP...1....4Y,2020MNRAS.492.3529S}. The growth rate and the nature of the unstable wave in the linear growing stage can be predicted by linear instability theory. Also, the linear instability can predict the stable parameter regime, which may correspond to plasma parameters in the nonlinear saturation stage. However, the linear theory cannot describe the development of the electron velocity distribution and unstable waves. The quasi-linear theory can explore the energy transfer between unstable waves and charged particles, and trace the dynamical evolution of the particle velocity distribution \citep[e.g.,][]{2017RvMPP...1....4Y}.

For the instability driven by the electron temperature anisotropy (e.g., the electromagnetic electron cyclotron instability and the periodic electron firehose instability), both linear and quasi-linear theories give the nearly the same predictions for the relaxation of the temperature anisotropy \citep{2016JGRA..121.9356S,2017PhPl...24a2907S,2017JGRA..122.4410K,2017PhPl...24k2104Y,2018JGRA..123....6L,2019ApJ...871..237S}.

For the whistler heat flux instability (the electron beam-driven whistler instability), \cite{2019MNRAS.486.4498S} analyzed the instability development at an initial plasma condition ($V_{\mathrm{eb}}=40V_{\mathrm{A}}$, and $A_{\mathrm{ec}}=A_{\mathrm{eb}}=1$; Case 3 in this reference) through the quasi-linear theory, and found the faster inhibition of the instability due to the induced temperature anisotropy of core and beam electrons. The electron beam speed in the saturation stage is larger than the threshold predicted by linear theory \citep{2019MNRAS.486.4498S}. \cite{2020MNRAS.492.3529S} further considered the development of the whistler heat flux induced by the interplay of the electron beam and the electron temperature anisotropy, and found that the temperature anisotropy in the saturation stage is lower than threshold predicted by linear theory \citep[also see][]{2019A&A...627A..76S,2019MNRAS.483.5642S}. \cite{2020JGRA..12527380S} used quasi-linear theory to explore the development of both forward and backward unstable whistler waves resulting from the electron beam and the electron anisotropic temperature, and also proposed that the saturation stage cannot be predicted by linear theory. Since the whistler heat flux instability is sensitive to the plasma parameters ($V_{\mathrm{eb}}$, $A_{\mathrm{ec}}$, $A_{\mathrm{eb}}$, $\beta_{\mathrm{ec\parallel}}$ and $\beta_{\mathrm{eb\parallel}}$), previous linear theories consider incomplete parameters, which may be one of reasons for discrepancy between linear and quasi-linear predictions. Furthermore, since the quasi-linear theory model is based on the growth rate (or damping rate) obtained from linear theory, the linear theory indeed predicts that the growth rate is totally reduced in the saturation stage \citep[see Figure 10 in][which presents the growth rate at several typical times in development of the whistler heat flux instability]{2019MNRAS.486.4498S}.

It should be noted that both quasi-linear theory and particle-in-cell simulation results proposed that parallel-propagating whistler waves induced by the heat flux instability cannot effectively scatter strahl electrons to halo electrons in the solar wind 
\citep{2019ApJ...882...81K,2019ApJ...882L...8L,2019MNRAS.486.4498S}. An alternate candidate is the oblique whistler wave \citep{2019ApJ...886..136V}. Besides oblique whistler waves driven by the electron beam with large flowing velocity in plasma with isotropic temperatures \citep{2019ApJ...886..136V}, this study proposes that these waves can be generated by the electron beam with small flowing velocity in plasma with anisotropic temperatures. To identify the effective interactions between oblique whistler waves and strahl electrons in the solar wind, it needs to study the development of oblique whistler waves under different plasma conditions through quasi-linear theory and simulations.

To summarize, our results propose that (1) the differential drift velocity among different electron populations in the solar wind may be constrained by the electron beam-driven acoustic/magnetoacoustic wave instability in low-$\beta_{\mathrm{ec\parallel}}$ regime, by the electron beam-driven whistler wave instability in the medium- and large-$\beta_{\mathrm{e\parallel}}$ regime, and by the oblique fast-magnetosonic/whistler instability in the regime of $\beta_{\mathrm{e\parallel}}\sim0.1-2$ and $A_{\mathrm{ec}}<1$; and (2) the electron temperature anisotropy in the solar wind is constrained by the electromagnetic electron cyclotron instability in the perpendicular temperature anisotropy regime, and by the aperiodic electron firehose instability and the ordinary-mode instability in the parallel temperature anisotropy regime.

\begin{acknowledgments}
This work was supported by the NNSFC 41531071, 41974203, 11673069, 11761131007, 11873018, 11790302.
\end{acknowledgments}

\section*{\textbf{Appendix A \\Instabilities driven by the electron temperature anisotropy}}

Figures \ref{fig:figA1} and \ref{fig:figA2} present the distributions of electron instabilities driven by the electron temperature anisotropy. The plasma parameters are the same as those used in Figures \ref{fig:fig1}$-$\ref{fig:fig8}, except both core and beam electron drift velocities are set to zero. Figure \ref{fig:figA1} and  \ref{fig:figA2}  exhibit the wave frequency and electromagnetic responses for unstable waves. The electromagnetic electron cyclotron instability generates whistler waves with $\omega_r\sim 0.1\omega_{\mathrm{ce}}$, $E_\parallel\lesssim0.7E$, $B_\parallel\lesssim B_\perp$, $|B_y|\simeq |B_x|$, and $\phi_{B_y-B_x}\simeq-90^\circ$. Unstable electron mirror-mode waves have $\omega_r=0$, $E_\parallel\gtrsim0.4E$, $B_\parallel\gtrsim B_\perp$, $|B_y|\sim 0.01-50|B_x|$, and $\phi_{B_y-B_x}\simeq0^\circ$. The periodic electron firehose instability generates the waves with $\omega_r\sim10^{-3}\omega_{\mathrm{ce}}$, $E_\parallel\lesssim0.7E$, $B_\parallel\sim 180$, $|B_y|\sim |B_x|$, and $\phi_{B_y-B_x}\simeq 90^\circ$. The aperiodic electron firehose instability produces zero-frequency mode waves with $E_\parallel\gtrsim0.5E$, $B_\parallel\lesssim 0.3B_\perp$, $|B_y|\gtrsim |B_x|$, and $\phi_{B_y-B_x}\simeq 0^\circ$. Besides, the unstable ordinary-mode waves have  $\omega_r\sim 0.7\omega_{\mathrm{ce}}$, $E_\parallel\sim E$, $B_\parallel\sim 0$, $|B_y|\gg |B_x|$, and $\phi_{B_y-B_x}\simeq-90^\circ$.

Figure \ref{fig:figA1} also exhibits the dominant  $A_{\mathrm{ec}}$ and $\beta_{\mathrm{ec\parallel}}$  region for each instability. The electromagnetic electron cyclotron instability and the electron mirror instability dominate the $A_{\mathrm{ec}}>1$ and $\beta_{\mathrm{ec\parallel}}\gtrsim 0.02$ region. The periodic and aperiodic electron firehose instabilities dominate the region with $A_{\mathrm{ec}}<1$ and $\beta_{\mathrm{ec\parallel}}\gtrsim 1.5$. The ordinary-mode instability controls the quasi-perpendicular instability in the region where $A_{\mathrm{ec}}<0.4$ and $\beta_{\mathrm{ec\parallel}}\simeq 1.3-3$.

Figure \ref{fig:figA2} shows the ($k,\theta)$ distributions for each instability. The electromagnetic electron cyclotron instability and the electron mirror instability distribute in the angle range of $\theta\lesssim35^\circ$ and $\theta\gtrsim35^\circ$, respectively. The periodic electron firehose instability dominates the small angle ($\theta\lesssim10^\circ$) region, and the aperiodic electron firehose instability arises at large angles ($\theta\simeq10^\circ-70^\circ$). The ordinary-mode instability are triggered in the range of $\theta\simeq88.5^\circ-89.6^\circ$. It should be noted that for the electromagnetic electron cyclotron, periodic electron firehose, and ordinary-mode instabilities, they produce forward and backward waves with the same instability distributions.

\section*{\textbf{Appendix B \\Instabilities driven by the electron beam}}

Figures \ref{fig:figA3} and \ref{fig:figA4} present electron instabilities resulting from the electron beam. The electron beam can induce the electron acoustic instability and the whistler instability. From the ($\beta_{\mathrm{ec\parallel}},V_{\mathrm{eb}}$) distributions, we can see that the electron beam-driven electron acoustic/magnetoacoustic instability dominates in the region with $V_{\mathrm{eb}}\gtrsim 15V_{\mathrm{A}}$ and $\beta_{\mathrm{ec\parallel}}\lesssim 0.1$, and the electron beam-driven whistler instability distributes in the region with $V_{\mathrm{eb}}\gtrsim 10V_{\mathrm{A}}$ and $\beta_{\mathrm{ec\parallel}}\gtrsim 0.1$. Moreover, for the former instability, the threshold value of the electron beam speed approximates $V_{\mathrm{eb}}\simeq 3V_{\mathrm{Tec}}$.

From the ($k,\theta$) instability distributions in Figure \ref{fig:figA4}, we see that the electron beam-driven electron acoustic/magnetoacoustic waves have $\theta\sim 0^\circ-89^\circ$, $\lambda_ek\sim10^2-10^3$, $\omega_r\sim 180 \omega_{\mathrm{ce}}\simeq\omega_{\mathrm{pe}}$, and $E_\parallel\simeq E$. When $\theta \neq0^\circ$, these waves have magnetic responses, $|B_y|>|B_x|$ and $\phi_{B_y-B_x}\simeq90^\circ$. At the low-$\beta_{\mathrm{ec\parallel}}$ condition, the electron beam also excites oblique whistler waves with $\theta\neq0^\circ$, and their electromagnetic responses are $E_\parallel\sim0.1E$, $|B_y|\sim|B_x|$ and $\phi_{B_y,B_x}\simeq90^\circ$. However, the electron beam-driven whistler instability ($\gamma\sim10^{-2}\omega_{\mathrm{ce}}$) is much weaker than the electron acoustic/magnetoacoustic instability ($\gamma\sim17\omega_{\mathrm{ce}}$). Note that \cite{2011PhPl...18h2902G} found that the electromagnetic electron cyclotron instability can drive oblique whistler waves in low-$\beta_{\mathrm{e\parallel}}$ ($\sim0.01$) plasma, which has the growth rate $\sim10^{-2}\omega_{\mathrm{ce}}$ much smaller than the corresponding values in the electron acoustic/magnetoacoustic instability. In the plasmas with $\beta_{\mathrm{ec\parallel}}\gtrsim 0.1$, electron beam-driven whistler waves ($\omega_r\simeq 0.03-0.1\omega_{\mathrm{ce}}$) propagate at small angles $\theta \simeq 0^\circ-10^\circ$, which is obviously different from unstable whistler waves in low-$\beta_{\mathrm{ec\parallel}}$ plasma.



\end{document}